\documentclass[showpacs,preprint,prb,amsmath,amssymb,floatfix,endfloats*]{revtex4}
\usepackage{graphicx}
\usepackage{epstopdf}
\usepackage{times}

\newcommand \op {\oplus}
\newcommand \om {\ominus}

\begin{document}

\title{Equilibrium and Stability of Polarization in Ultrathin Ferroelectric Films\\ 
with Ionic Surface Compensation} 

\author{G. Brian Stephenson}
\altaffiliation{Electronic mail: stephenson@anl.gov}
\author{Matthew J. Highland}
\affiliation{Materials Science Division, 
Argonne National Laboratory, Argonne, Illinois 60439}

\date{\today}

\begin{abstract}  
Thermodynamic theory is developed 
for the ferroelectric phase transition of an ultrathin film
in equilibrium with a chemical environment that supplies
ionic species to compensate its surface.
Equations of state and free energy expressions are developed
based on Landau-Ginzburg-Devonshire theory,
using electrochemical equilibria to provide
ionic compensation boundary conditions.
Calculations are presented for a monodomain PbTiO$_3$ (001) film
coherently strained to SrTiO$_3$
with its exposed surface and its electronically conducting bottom electrode
in equilibrium with a controlled oxygen partial pressure.
The stability and metastability boundaries of phases of different polarization are determined
as a function of temperature, oxygen partial pressure, and film thickness.
Phase diagrams showing polarization and internal electric field are presented.
At temperatures below a thickness-dependent Curie point,
high or low oxygen partial pressure stabilizes positive or negative polarization, respectively.
Results are compared to the standard cases of electronic compensation
controlled by either an applied voltage or charge across two electrodes.
Ionic surface compensation through chemical equilibrium with an environment
introduces new features into the phase diagram.
In ultrathin films, a stable nonpolar phase can occur between
the positive and negative polar phases when varying the external chemical potential 
at fixed temperature,
under conditions where charged surface species are not present in sufficient
concentration to stabilize a polar phase.
\end{abstract}

%77.80.bn Ferroelectricity and antiferroelectricity, Phase transitions and Curie point, strain and interface effects
%64.70.Nd Structural transitions in nanoscale materials
%68.55.-a Thin film structure and morphology
%68.43.-h Chemisorption/physisorption: adsorbates on surfaces
%77.84.Cg Ferroelectric materials, PZT and other titanates

\maketitle
\vskip 1 pc

\section{Introduction}

The equilibrium polarization structure of an ultrathin ferroelectric film is strongly affected
by the nature of the charge compensation of its interfaces.
When there is insufficient free charge at the interfaces,
a locally polar state can be stabilized by 
formation of equilibrium 180$^\circ$ stripe domains
that reduce the depolarizing field 
energy.\cite{02_StriefferPRL_89_067601,04_Fong_Science,GBS06JAP,LAI07APL,CT08APL,BRAT08JCTN}
When electrodes are present,
electronic charge at the interfaces can stabilize a monodomain polar state,
provided that the effective screening length in the electrodes is sufficiently small
compared with the film 
thickness.\cite{BRAT08JCTN,BAT73PRL, BAT73JVST,WURF76FERRO,DAWB03JPCM,
TAG06JAP,JUNQ03NAT, SAI05PRB,09_Stengel_NatMat}
In both cases, the Curie point $T_C$ is expected to be increasingly suppressed
as film thickness decreases because of the residual depolarizing field energy.

Even when the surface electrode is missing,
it has been experimentally observed that
a monodomain polar state can be stable in ultrathin ferroelectric 
films.\cite{LICHT05PRL,FONG06PRL,06_DespontPRB_73_094110,LICHT07APL}
This has been attributed to the presence of 
ionic species at the surface that provide charge compensation
and reduce the depolarizing field energy,\cite{FONG06PRL}
similar to the adsorbates observed on bulk ferroelectric surfaces.\cite{06_KalininNL_4_555}
Furthermore, experiments have shown that the sign of the polarization can be inverted
by changing the chemical environment in equilibrium with the 
surface.\cite{09_WangPRL_102_047601,10_Kim_APL96_202902}
Recently\cite{10_HighlandPRL} it was found that when the polarization
is inverted by changing the external chemical potential,
switching can occur without domain formation
and at an internal field reaching the intrinsic coercive field
for certain ranges of film thickness and temperature.
Thus, through either kinetic suppression of domain nucleation,
or the structure of the equilibrium phase diagram,
an instability point of the initial polar state can be reached.
This is in sharp contrast to switching by applied field across electrodes,
where the consensus has been that polarization inversion occurs
only by domain nucleation and growth
at fields well below the instability.\cite{05_Dawber_RevModPhys77_1083}

These studies motivate the need to understand the polarization phase diagrams
and metastability limits for ultrathin ferroelectric films with ionic surface compensation,
in chemical equilibrium with their environment.
While the energy and structure of ferroelectric surfaces compensated by ions have been
predicted by {\it ab initio} 
calculations,\cite{FONG06PRL,09_WangPRL_102_047601,06_Spanier_NanoLett6_735,09_ShinNL_9_3720}
these zero-temperature results have been extrapolated to experimental temperatures
using simple entropy estimates,
and to date have not included the effects of interaction with the ferroelectric phase transition
and $T_C$.
Here we develop a thermodynamic theory
for the ferroelectric phase transition of an ultrathin film
in an environment that supplies
ionic species to compensate the polarization discontinuity 
at the surface of the ferroelectric.
We use an approach based on Landau-Ginzburg-Devonshire (LGD) theory
for the ferroelectric material,\cite{GBS06JAP}
with boundary conditions that include 
both the depolarizing field effects that arise in ultrathin films
and the creation of ionic surface charge through electrochemical equilibria.
This new chemical boundary condition is based on a
Langmuir adsorption isotherm for ions.\cite{96_Schmickler}
We develop an expression for the free energy of the system
and use it to determine 
the equilibrium monodomain polarization states and their stability.
For simplicity we do not include additional ``intrinsic" surface effects
or polarization gradients in the ferroelectric.\cite{Kretschmer}
We compare and contrast  
our model for ionic compensating charge
controlled by an applied chemical potential
with existing models for electronic compensating charge
controlled by either an applied voltage or fixed charge,
to elucidate how the present predictions for ionic compensation 
differ from prior work.

We find that the equilibrium phase diagram of a monodomain ferroelectric film
as a function of temperature and chemical potential
can have a different form than the standard phase diagrams 
as a function of temperature and applied voltage or charge.
We present calculations for PbTiO$_3$ (001) films
with a conducting bottom electrode (e.g., SrRuO$_3$),
coherently strained to SrTiO$_3$,
and with a surface compensated 
by excess or missing 
oxygen ions.\cite{FONG06PRL,09_WangPRL_102_047601}
For sufficiently thin films, we find that a nonpolar state becomes stable
between the positive and negative polar states,
within the range of external oxygen partial pressures
where there is insufficient surface charge
to stabilize a polar state.
Under these conditions the Curie temperature 
depends strongly on the oxygen chemical potential.

\section{Thermodynamic Model}

In this section we establish the electrostatic boundary condition,
the ferroelectric constitutive relation,
and the free energy expressions
used to describe an ultrathin ferroelectric film.
We consider a uniformly polarized, monodomain film with uniaxial polarization 
oriented out-of-plane (normal to the interfaces).
This should apply to systems 
such as PbTiO$_3$ (001) coherently strained to SrTiO$_3$,
since LGD theory\cite{GBS06JAP,01_Koukhar_PRB64_214103} predicts that
compressive in-plane strain favors this ``$c$ domain'' polarization orientation.
Even if out-of-plane polarization is suppressed by depolarization field effects,
in this system the in-plane ``$a$ domain'' polarization orientation 
is less stable than the nonpolar phase\cite{10_HighlandPRL}
at temperatures above 360~K.
For this case all fields can be specified by scalars 
since their in-plane components are zero.

\begin{figure}
\centering
\includegraphics[width=3.0in]{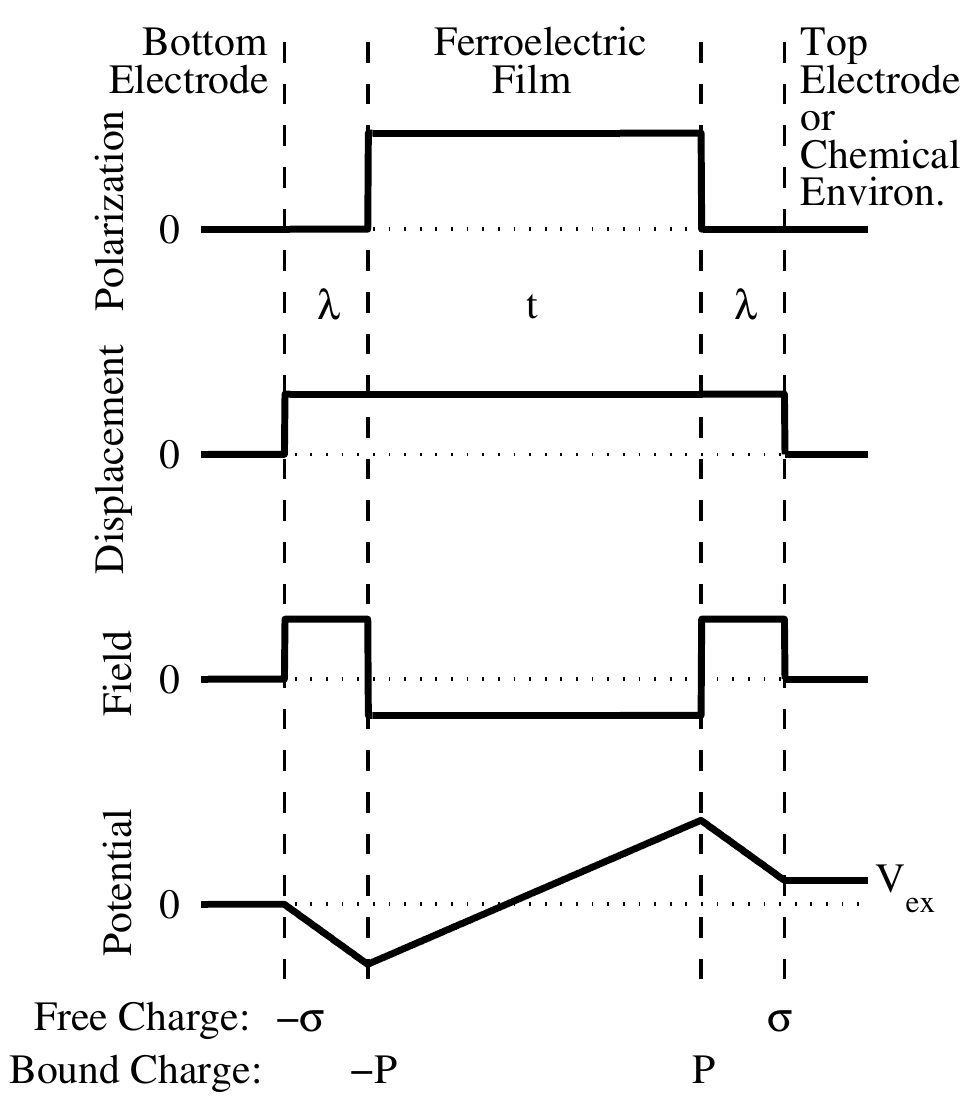} 
\caption{Schematic of polarization, displacement, 
electric field, and electric potential 
in the bulk and at the interfaces of a ferroelectric film 
of thickness $t$ and polarization $P$.
Compensating planes of charge density $\sigma$ 
can be considered to reside at a separation $\lambda$
equal to the effective screening length in the electrodes.}
% (chem\_ferroN2new\_1.pdf)}
\label{F1}
\vspace{-0.1in}
\end{figure}

To include the effects of incompletely neutralized depolarizing field,
we use the simple electrostatic model illustrated in Fig.~\ref{F1}.
The spatial separation between the compensating free charge in the electrodes
and the bound charge at the outer surfaces of the ferroelectric
leads to residual depolarizing field in the film 
even when the external voltage $V_{ex}$ is zero (i.e. short-circuit conditions).
Figure~\ref{F1} shows the polarization $P$ and
displacement $D$ in a ferroelectric film of thickness $t$
sandwiched on the top and bottom 
by planes of compensating free charge of density $\pm \sigma$,
at a distance $\lambda \ge 0$ outside the ferroelectric.
The planes of bound and free charge lead to steps in $P$ and $D$, 
respectively.
In Fig.~\ref{F1},
$P$ and $D$ are positive 
and the free charge on the top electrode $\sigma$ is negative.
The electric field and potential can be calculated from 
$E = (D - P)/\epsilon_0$ and $\nabla \phi = -E$,
where $\epsilon_0$
is the permittivity of free space.
The internal field in the ferroelectric film is
$E_{in} = -(\sigma + P)/\epsilon_0$,
while the field just outside the film is
$E_{\lambda} = -\sigma/\epsilon_0$.

In a series of early papers,
Batra, Wurfel, and Silverman\cite{BAT73PRL, BAT73JVST,WURF76FERRO}
showed that the results of a more complex model taking into account
the space charge distribution and nonzero screening length 
in non-ideal metal electrodes
could be reproduced by this simple model in which ideal metal electrodes
are separated from the ferroelectric by a vacuum gap of width equal to
the screening length, and all bound and free charges reside at the interfaces.
This model has been discussed 
extensively\cite{BRAT08JCTN,DAWB03JPCM, LICHT05PRL, TAG06JAP}
and used to parametrize the results of {\it ab initio} 
calculations.\cite{JUNQ03NAT}
An alternative description in terms of interfacial 
capacitance\cite{09_Stengel_NatMat,06_Stengel_Nature}
is equivalent if the interfacial capacitance per unit area 
is identified with $\epsilon_0/\lambda$.
Recent 
calculations\cite{SAI05PRB,09_Stengel_NatMat,09_StengelPRB_80_224110} 
have shown that
the screening length for the electrode material can be generalized to be
an effective screening length for a given ferroelectric/electrode interface.

% figure 1: schematic of surface charges, polarization, field, and potential

We can relate the external voltage $V_{ex}$ across the structure
to $\sigma$ and $P$
by integrating the field to give
\begin{equation}
\epsilon_0 V_{ex} = 2 \lambda \sigma + t (\sigma + P).
\label{eq3}
\end{equation}
The field in the ferroelectric can then be expressed 
as a function of $P$ and either $\sigma$ or $V_{ex}$ using
\begin{eqnarray}
E_{in} &=& -(\sigma + P)/\epsilon_0 \nonumber \\*
&=& - \frac{V_{ex} + 2 \lambda P/\epsilon_0}{2\lambda + t}.
\label{eq4}
\end{eqnarray}
In the latter expression, the second term in the numerator gives the 
voltage from the residual depolarizing field that 
is proportional to (and opposes) the film polarization.

In this simple electrostatic model, we assume that the two interfaces 
have the same screening length and work function. 
In a polarized material, 
these quantities can depend upon the polarization magnitude and orientation
with respect to the surface, and differences between the two interfaces may arise
even if the electrode materials are identical.\cite{SAI05PRB,09_StengelPRB_80_224110}
Since to first order these effects simply
add a term to $2 \lambda$ in the numerator of Eq.~(\ref{eq4}),
which is already a variable parameter in our model,
we have neglected these differences.
The approximations in this electrostatic model
are not critical in determining the new behaviors we find
below for ionic surface compensation 
(which occur even for $\lambda = 0$),
but do provide simple, analogous electronic compensation models for comparison.

To determine the equilibrium polarization in the film,
the values of field and polarization inside the ferroelectric 
must simultaneously satisfy both the electrostatic boundary condition Eq.~(\ref{eq4})
and the $E_{in}(P)$ constitutive relation for the ferroelectric. 
For PbTiO$_3$ this can be written as\cite{GBS06JAP}
\begin{equation}
E_{in} = f^{\prime}(P) = 2\alpha_3^* P + 4\alpha_{33}^* P^3 + 6\alpha_{111} P^5,
\label{eq5}
\end{equation}
where $f^{\prime}(P)$ is the derivative of the bulk LGD
free energy density
\begin{equation}
f(P) \equiv \alpha_3^* P^2 + \alpha_{33}^* P^4 + \alpha_{111} P^6,
\label{eq5b}
\end{equation}
and the coefficients $\alpha_i^*$ are those 
for a coherently-strained film,\cite{98_PertsevPRL_80_1988}
\begin{eqnarray}
\alpha_3^* &=&
\frac{T - T_0}{2\epsilon_0 C} - \frac{2 x_m Q_{12}}{s_{11} + s_{12}}, \nonumber \\*
\alpha_{33}^* &=& \alpha_{11} + \frac{Q_{12}^2}{s_{11} + s_{12}},
\end{eqnarray}
where $x_m$ is the epitaxial misfit strain of the zero polarization state,
$T_0$ is the temperature at which $\alpha_3^*$ changes sign for $x_m = 0$,
$C$ is the Curie constant,
 and $Q_{ij}$ and $s_{ij}$ are the electrostrictive and elastic compliance coefficients,
respectively.\cite{GBS06JAP,HAUN87JAP,ROS98JMR}
Values of these material parameters for PbTiO$_3$ are listed in Table~\ref{tab1}.
The misfit strain $x_m$ 
has  a somewhat temperature-dependent value\cite{GBS06JAP} 
of about -0.01 for PbTiO$_3$ coherently strained to SrTiO$_3$.
While for unstressed bulk PbTiO$_3$ the fourth-order polarization coefficient $\alpha_{11}$ 
is slightly negative,
indicating a weakly first-order transition as a function of $T$ at $E_{in} = 0$,
for coherently strained films the coefficient $\alpha_{33}^*$ has a positive value of 
$5.0 \times 10^7$ Vm$^5$/C$^3$,
indicating that the transition is second order.\cite{98_PertsevPRL_80_1988}

The strain normal to the film can be calculated from the polarization 
using the expression\cite{GBS06JAP}
\begin{equation}
x_3 = Q_{11} P^2 + 2 s_{12}(x_m - Q_{12} P^2)/(s_{11} + s_{12}).
\label{eq13}
\end{equation}
If the effects of depolarizing field are neglected
(i.e. for ideal electrodes with $\lambda = 0$),
the Curie temperature $T_C^{\circ}$ is determined 
by the change in sign of $\alpha_{3}^*$,
which gives
\begin{equation}
T_C^{\circ} = T_0 + 4 \epsilon_0 C Q_{12} x_m / (s_{11} + s_{12}).
\end{equation}
Using the $x_m(T)$ appropriate for epitaxially strained PbTiO$_3$ on SrTiO$_3$,
this gives $T_C^{\circ} = 1023$~K, 
about 270~K higher than in the $x_m = 0$ case.

\begin{table}
\caption{Material parameters\cite{GBS06JAP,HAUN87JAP,ROS98JMR} for PbTiO$_3$.}
\label{tab1}
\begin{ruledtabular}
\begin{tabular}{cccccc}

$T_0$ &752.0 &(K) 
&$Q_{11}$ &$8.9 \times 10^{-2}$ &(m$^4$/C$^2$) \\ 

$C$ &$1.5 \times 10^{5}$ &(K)
&$Q_{12}$ &$-2.6 \times 10^{-2}$ &(m$^4$/C$^2$) \\ 

$\alpha_{11}$ &$-7.25 \times 10^{7}$ &(Vm$^5$/C$^3$) 
&$s_{11}$ &$8.0 \times 10^{-12}$ &(m$^2$/N) \\ 

$\alpha_{111}$ &$2.61 \times 10^{8}$ &(Vm$^9$/C$^5$)
&$s_{12}$ &$-2.5 \times 10^{-12}$ &(m$^2$/N) \\ 

\end{tabular}
\end{ruledtabular}
\end{table}

We can determine which of the equilibrium solutions is stable, metastable, or unstable
by considering the appropriate free energy of the system.
For a closed system (e.g. fixed charge),
the Helmholtz free energy is minimized at equilibrium.
The Helmholtz free energy per unit area ${\cal A}$ 
can be written as\cite{GBS06JAP}
\begin{eqnarray}
{\cal A}  &=& t[f(P) +\frac{ \epsilon_0}{2} E_{in}^2 ] 
+ 2\lambda \frac{ \epsilon_0}{2} E_{\lambda}^2
\nonumber \\*
&=&  t \left [ f(P) +\frac{ (\sigma + P)^2}{2\epsilon_0} \right ] 
+ \frac{ \lambda \sigma^2}{\epsilon_0}, 
\label{eq5a}
\end{eqnarray}
where the two terms are for the ferroelectric film and surrounding screening regions.
For an open system (e.g. fixed potential),
the Gibbs free energy is minimized at equilibrium.
The Gibbs free energy per unit area ${\cal G}$ 
is given by\cite{GBS06JAP}
\begin{equation}
{\cal G}  = {\cal A} - V_{ex} \sigma,
\label{eq5c}
\end{equation}
where the difference between the Gibbs and Helmholtz free energies
$-V_{ex} \sigma$
is the electrical work done on the system by the external circuit.
This difference is in accord with that 
in a recent derivation\cite{09_StengelNatPhys_5_304} 
of the energy functionals
to be minimized in first-principles calculations at fixed $D$ and fixed $E$.

\section{Ferroelectric film with electronic compensation}

In this section we present the equations of state and phase diagrams
for ferroelectric films with electronic compensation
under controlled voltage or charge conditions,
as background for development of theory for ionic compensation.
Some of the more subtle differences between fixed voltage and fixed charge
boundary conditions are highlighted.

% figure 2: solving for P and E
\begin{figure}
\centering
\includegraphics[width=3.0in]{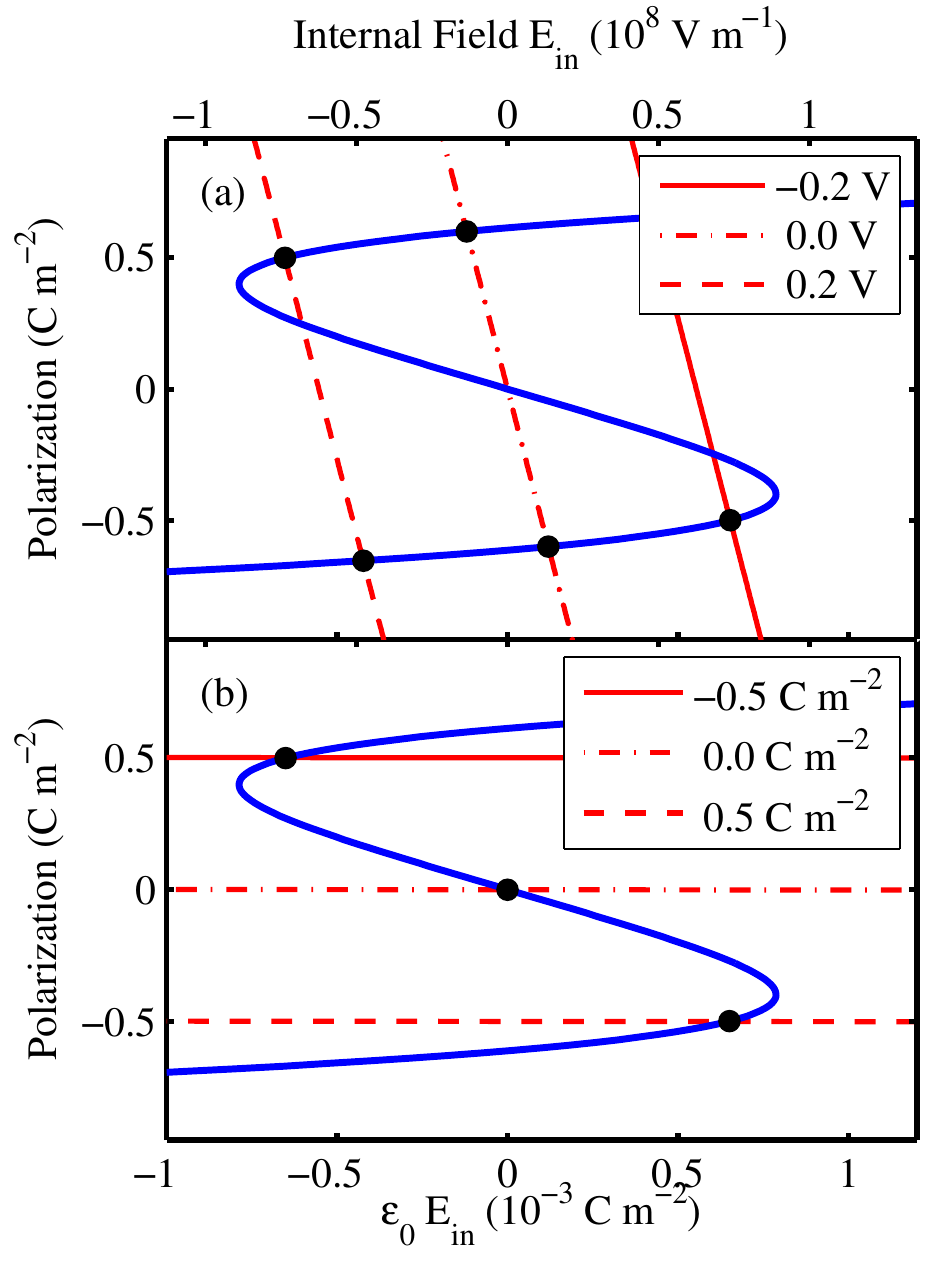} 
\caption{(Color online) Polarization vs. internal field.
Straight red lines  show 
(a) fixed $V_{ex}$ (at $t$ = 3.2~nm and $\lambda/t = 10^{-4}$) or 
(b) fixed $\sigma$ boundary conditions 
from Eq.~(\ref{eq4}),
for the three values of the fixed quantity given in the legend.
In each case, the ``S'' shaped blue curve is the constitutive relation,
Eq.~(\ref{eq5}), for PbTiO$_3$ coherently strained to SrTiO$_3$ at 644~K.
Marked intersections correspond to stable or metastable equilibrium solutions.}
%(chem\_ferroN2\_2.pdf)}
\label{F2}
\vspace{-0.1in}
\end{figure}

Figure~\ref{F2} graphically shows the equilibrium polarization and field values
that simultaneously satisfy the constitutive relation, Eq.~(\ref{eq5}),
and either the fixed $V_{ex}$ or the fixed $\sigma$ boundary condition.
A temperature of 644~K was chosen to match one of the experimental conditions
previously studied.\cite{09_WangPRL_102_047601,10_HighlandPRL}
Each line in Fig.~\ref{F2}(a)
is the fixed $V_{ex}$ boundary condition
from the second equality of Eq.~(\ref{eq4})
for a particular value of $V_{ex}$.
The deviation of this line from vertical reflects the
nonzero value of $\lambda/t = 10^{-4}$
used to model the screening length in the electrodes.
As $V_{ex}$ is varied,
this boundary condition translates along the horizontal $E_{in}$ axis.
For $|V_{ex}|$ less than a certain value, there are three intersections 
representing equilibrium solutions;
at larger $|V_{ex}|$, there is only a single solution.
The marked intersections correspond to solutions
that are not unstable,
as described below.
Each line in Fig.~\ref{F2}(b)
is the fixed $\sigma$ boundary condition
from the first equality of Eq.~(\ref{eq4})
for a particular value of $\sigma$.
These lines are nearly horizontal,
showing that the field dependence of $P$
at constant $\sigma$ is negligible. 
As $\sigma$ is varied,
this boundary condition translates along the vertical $P$ axis.
In this  fixed, uniform $\sigma$ case,
there is a single equilibrium solution
at all $\sigma$ and $T$ values.
The behavior is independent of $\lambda$ and $t$,
and, as described below,
the equilibrium solution is always stable.

\subsection{Phase diagram for controlled $V_{ex}$}

\begin{figure}
\centering
\includegraphics[width=3.0in]{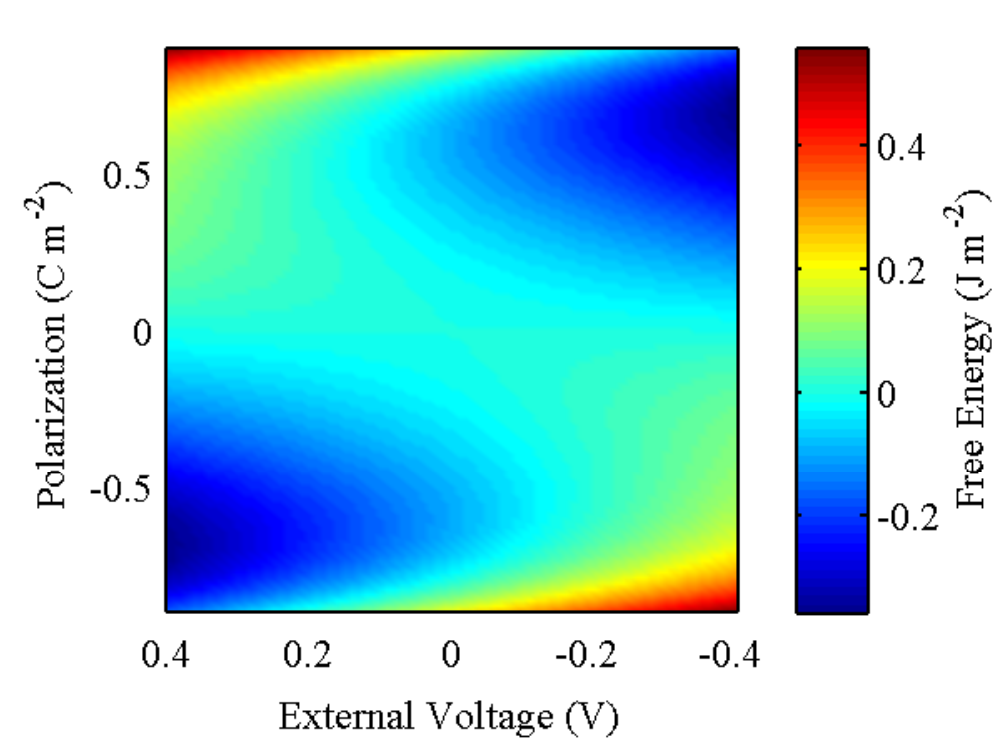} 
\caption{(Color online) Gibbs free energy ${\cal G}$ vs. $P$ and $V_{ex}$ for a
$t = 3.2$~nm PbTiO$_3$ film coherently strained to SrTiO$_3$ at 644~K, 
with  $\lambda/t = 10^{-4}$.
Color scale gives values of ${\cal G}$.}
% (chem\_ferroN2\_5.pdf)}
\label{F2aa}
\vspace{-0.1in}
\end{figure}

At constant $V_{ex}$,
the equilibrium polarization value is that which minimizes ${\cal G}$.
Using Eq.~(\ref{eq3}) to eliminate $\sigma$ gives an expression
for ${\cal G}$  in terms of
$V_{ex}$ and $P$,
\begin{equation}
{\cal G}  =  t f(P)
- \frac{\epsilon_0 V_{ex}^2}{2 (2\lambda + t)} 
+ \frac{t P V_{ex}}{2\lambda + t} 
+ \frac{\lambda t P^2}{\epsilon_0(2\lambda + t)}.
\label{eq5f}
\end{equation}
Figure~\ref{F2aa}  shows ${\cal G}$ as a function of $P$ and $V_{ex}$
corresponding to Fig.~\ref{F2}(a).
The equilibrium polarization $P_{eq}$ can be determined by
setting the first derivative of ${\cal G}$ at constant $V_{ex}$  to zero,
giving the equation of state
\begin{equation}
0 = \frac{1}{t} \left . \frac{\partial \cal G}{\partial P} \right | _ {V_{ex}}
= f^\prime (P) + \frac{V_{ex} + 2\lambda P /\epsilon_0}{2\lambda + t}.
\label{eq5ff}
\end{equation}
This agrees with the simultaneous solution of the constitutive relation
and boundary condition shown above,
Eqs.~(\ref{eq4}) and (\ref{eq5}).

% figure 2aa: G(P, V)

The stability of the equilibrium solutions of Eq.~(\ref{eq5ff})
is determined by the sign of the second derivative of ${\cal G}$ 
at that value of $P$,
\begin{equation}
\frac{1}{t} \left . \frac{\partial^2 \cal G}{\partial P^2} \right | _ {V_{ex}}
= f^{\prime\prime} (P) + \frac{2\lambda}{\epsilon_0(2\lambda + t)}.
\label{eq5g}
\end{equation}
When this is negative, the solution is unstable;
when it is positive, the solution is stable or metastable.
In particular, when there are three solutions, 
as shown in Fig.~\ref{F2}(a),
the middle one near $P = 0$ is unstable. 
The values of $P$ and $V_{ex}$ at the instability
(limit of metastability)
are given by the condition that both
the first and second derivatives of ${\cal G}$ are zero.
At this value of $P$,
the $P(E_{in})$ curve of the constitutive equation, Eq.~(\ref{eq5}), 
has the same slope as the constant $V_{ex}$ boundary condition line, 
Eq.~(\ref{eq4}),
in Fig.~\ref{F2}(a).
The value of $E_{in}$ at the instability 
is the intrinsic coercive field for the film/electrode system
with parameters $t$ and $\lambda$,
taking into account the effect of depolarizing field.

The solution of Eq.~(\ref{eq5g}) for $P=0$ gives the condition for
the Curie temperature $T_C$
which can differ from the value $T_C^{\circ}$ for $\lambda = 0$.
The change in $T_C$ due to a nonzero screening length 
is given by\cite{BAT73JVST}
\begin{equation}
\Delta T \equiv T_C - T_C^{\circ} =  \frac{-2\lambda C}{2\lambda + t}.
\label{eq5m}
\end{equation}
Because the Curie constant $C$ is much larger than $T_C^{\circ}$ 
for typical ferroelectrics,
stability of the polar phase requires $\lambda << t$.
Even a ratio $\lambda/t = 0.001$ gives $\Delta T = -300$~K for PbTiO$_3$.
Effective screening lengths $\lambda$ calculated 
from first principles\cite{JUNQ03NAT,09_Stengel_NatMat}
vary between zero and 0.02~nm for various electrode-ferroelectric interfaces.

% figure 2a: Polarization, strain, and free energy as a function of $V_{ex}$
\begin{figure}
\centering
\includegraphics[width=3.0in]{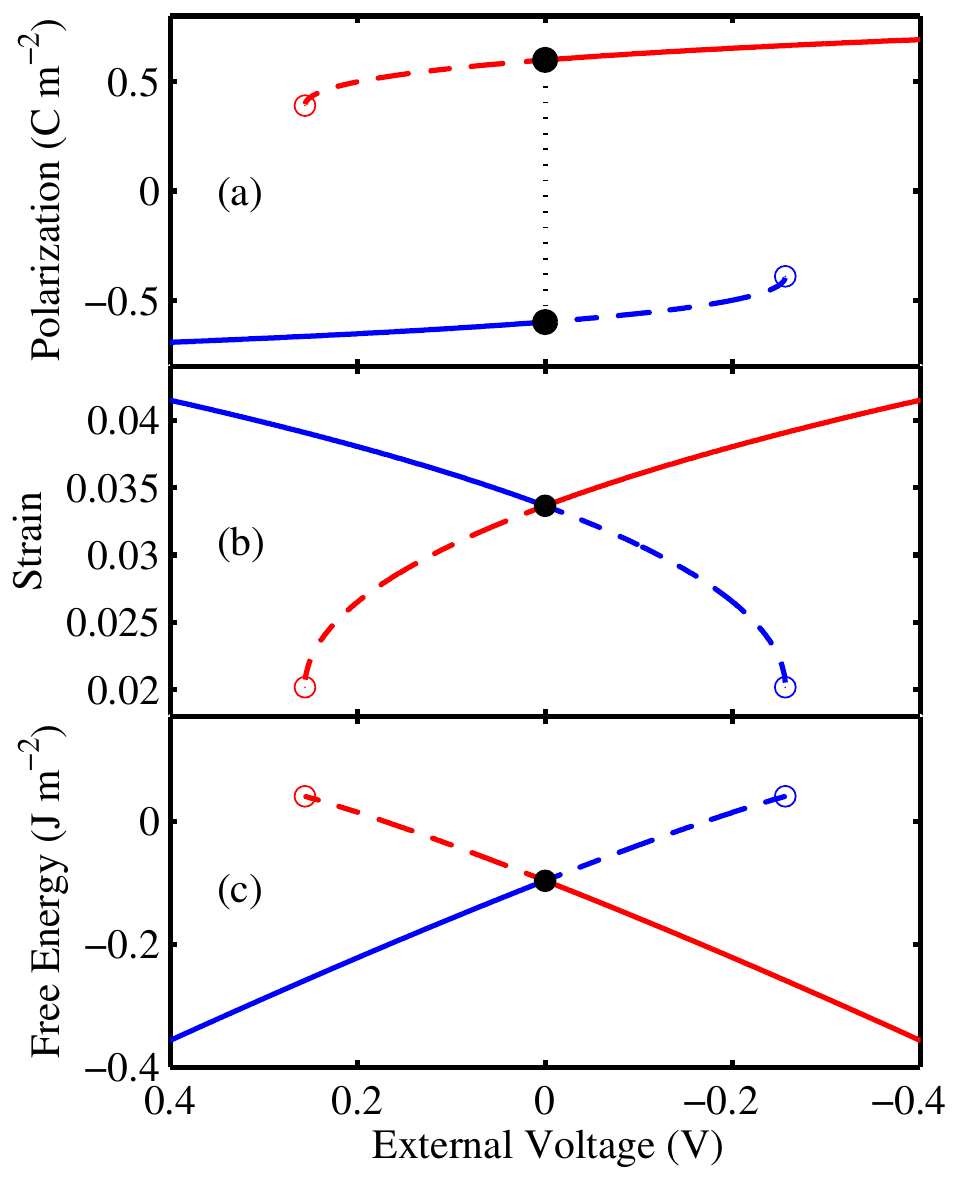} 
\caption{(Color online) Equilibrium solutions for (a) polarization $P$, 
(b) strain $x_3$, 
and (c) Gibbs free energy ${\cal G}$
as a function of $V_{ex}$,
calculated for PbTiO$_3$ coherently strained to SrTiO$_3$ at 644~K
with $t = 3.2$~nm and $\lambda/t = 10^{-4}$. 
Red (blue) curves are positive (negative) polarization; 
solid (dashed) segments are stable (metastable);
closed (open) circles show equilibrium transition (instability) points.}
%(chem\_ferroN2\_3.pdf)}
\label{F2a}
\vspace{-0.1in}
\end{figure}

% figure 2e: Polarization phase diagram as a function of $V_{ex}$ and T
\begin{figure}
\centering
\includegraphics[width=3.0in]{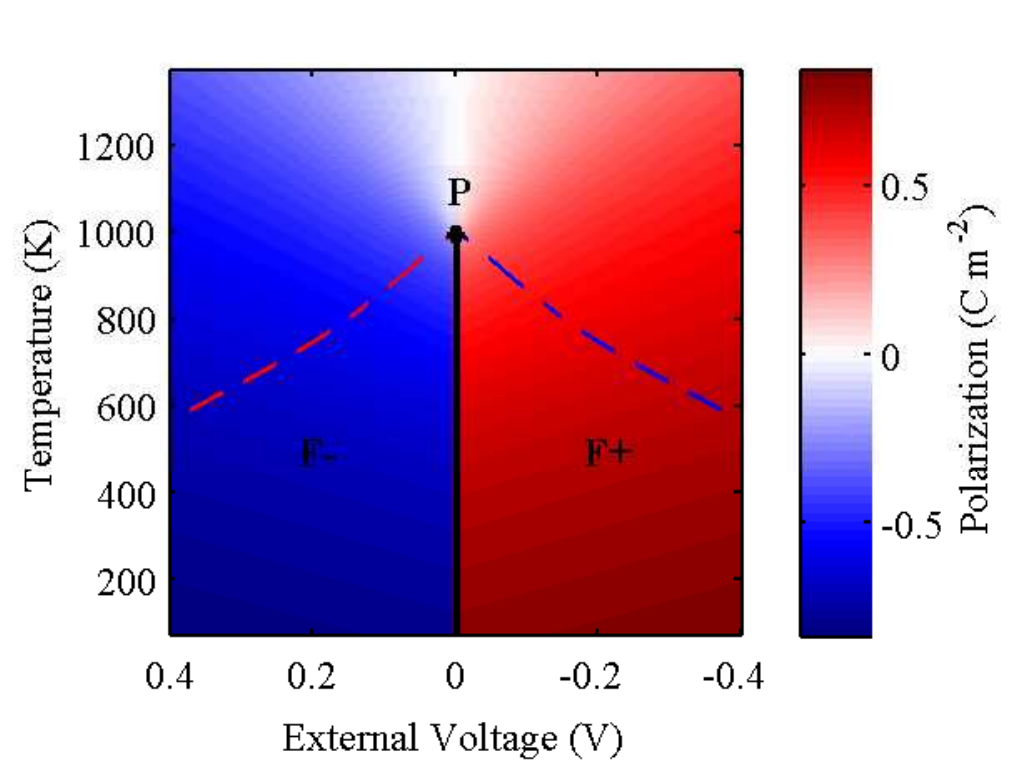} 
\caption{(Color online) Equilibrium polarization phase diagram 
as a function of $V_{ex}$ and $T$
for PbTiO$_3$ coherently strained to SrTiO$_3$ 
with $t = 3.2$~nm and $\lambda/t = 10^{-4}$.
Color scale gives polarization of stable phase.
Solid black line is phase boundary between 
positive and negative polar ferroelectric (F+ and F-) phases,
terminating at $T_C$. 
Dashed red and blue curves are metastability limits 
of the F+ and F- phases, respectively.}
%(chem\_ferroNX6\_3.pdf)}
\label{F2e}
\vspace{-0.1in}
\end{figure}

% figure 2f: Field phase diagram as a function of $V_{ex}$ and T
\begin{figure}
\centering
\includegraphics[width=3.0in]{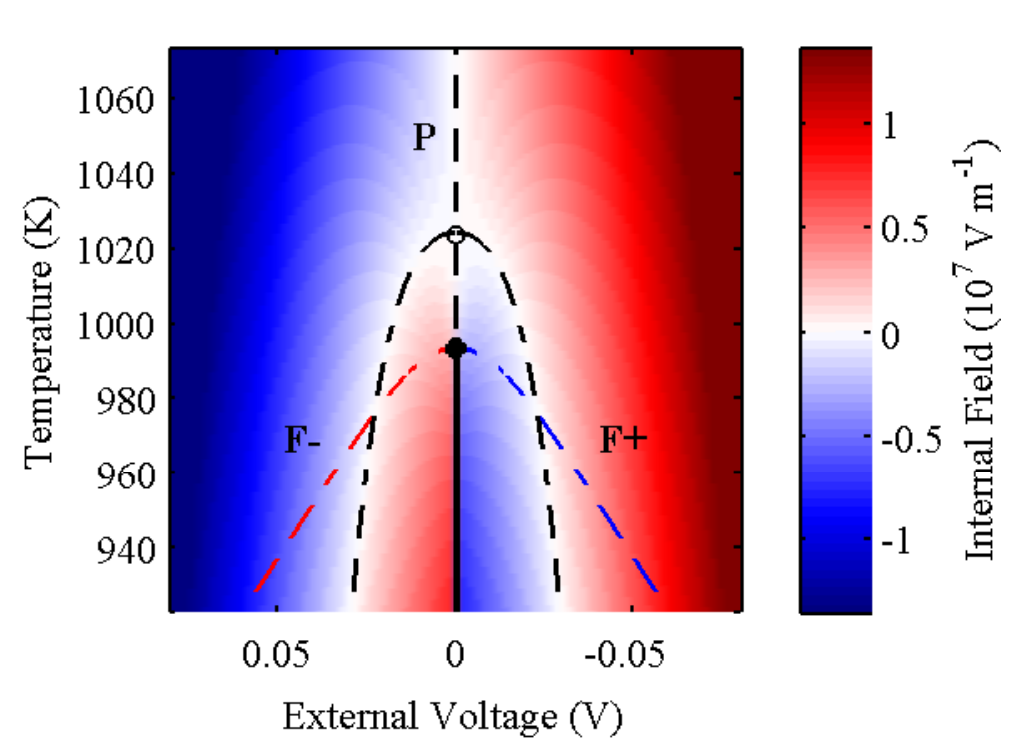} 
\caption{(Color online) Internal field
in region near the Curie point,
corresponding to Fig.~\ref{F2e}.
Color scale gives electric field in stable phase.
Dashed black curves show conditions for zero field,
which intersect at $T_C^{\circ}$ (open circle),
the Curie point for a film without depolarizing field $(\lambda = 0)$.}
%(chem\_ferroNX7\_5.pdf)}
\label{F2f}
\vspace{-0.1in}
\end{figure}

Figure \ref{F2a} shows the equilibrium polarization, strain, and  free energy 
of the stable and metastable solutions as a function of $V_{ex}$.
These indicate the possible polarization hysteresis and strain butterfly loops.
Two equilibrium solutions (one stable and one metastable) corresponding to 
oppositely polarized states
exist when $|V_{ex}|$ is smaller than the instability.
The stable solution switches between positive and negative polarization
at $V_{ex} = 0$.
At values of $V_{ex}$ in the metastable region between the equilibrium and instability points,
polarization switching requires nucleation of domains of the opposite polarity.
The nucleation barrier becomes zero when $V_{ex}$ reaches the instability.\cite{CAHN59JCP}
At values of $V_{ex}$ exceeding the instability,
switching occurs by a continuous process without nucleation.

Figure~\ref{F2e} shows the equilibrium polarization phase diagram 
as a function of $V_{ex}$ and $T$.
While there is a first-order transition phase between 
positive and negative polar ferroelectric (F+ and F-) phases,
this terminates at $T_C$ in a second-order transition to the nonpolar 
paraelectric (P) phase\cite{Strukov98}.
When $T$ is varied at nonzero values of $V_{ex}$, there is no phase transition between
the nonpolar and the stable polar phase.
The dashed red and blue curves are the limits
of the metastable F+ and F- phases, respectively.
The polarity switching transition driven by changing $V_{ex}$ at fixed $T$
requires nucleation under conditions inside (below) these curves,
and is continuous (non-nucleated) outside (above) these curves.

The nonzero screening length $\lambda$ not only depresses $T_C$ below $T_C^{\circ}$,
but also produces inverted electric fields in the film 
in the region near the phase boundary (small $|V_{ex}|$).
Figure~\ref{F2f} shows the internal field as a function of $V_{ex}$ and $T$
in the vicinity of $T_C$.
The inverted electric fields extend into the nonpolar phase
in the region between $T_C$ and $T_C^{\circ}$.
Thus, when a small external voltage is applied to a film in this region,
the equilibrium field in the film is {\it opposite} to the applied field.
Close to $T_C$, the magnitude of this inverted field
is larger than that of the applied field,
producing an (internal) voltage gain in a passive device
that diverges as $T_C$ is approached.
It has been proposed that such ``negative capacitance'' 
could be used to improve the performance of nanoscale 
transistors.\cite{Salahuddin08NL}
The conditions for zero field, shown as black dashed curves in Fig.~\ref{F2f},
can be obtained from Eq.~(\ref{eq4}) as
$V_{ex} = -(2\lambda/\epsilon_0)P_0$,
where $P_0$ is the zero-field spontaneous polarization 
of the epitaxially strained film.\cite{GBS06JAP}
Unlike $T_C$,
these boundaries are independent of film thickness.

% figure 6aa: A(P, D)
\begin{figure}
\centering
\includegraphics[width=3.0in]{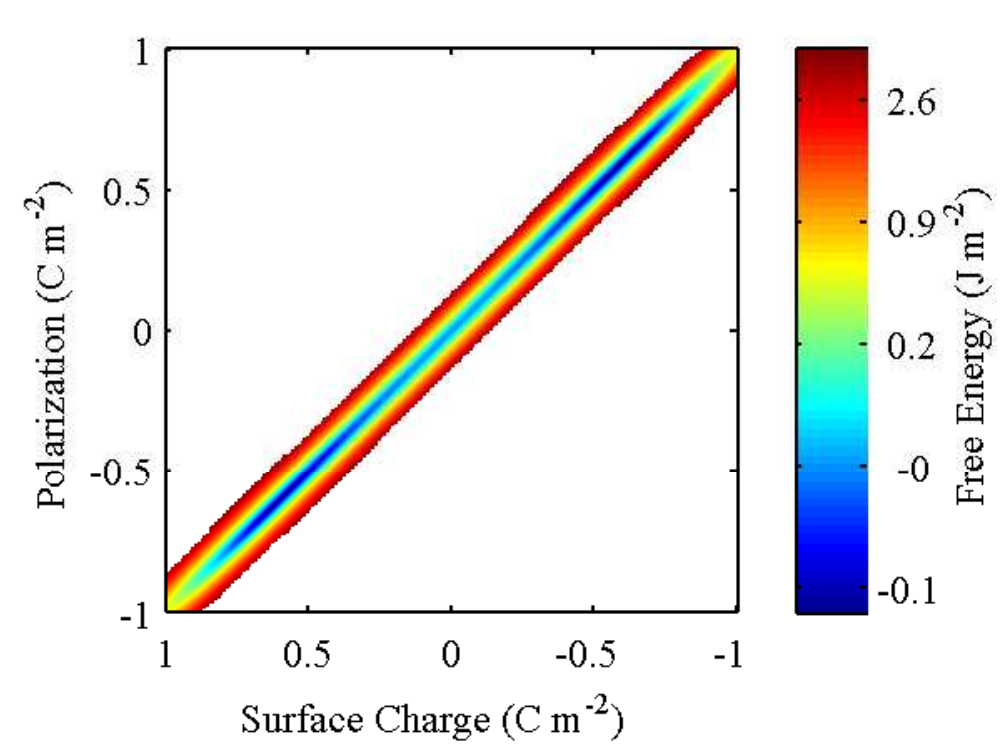} 
\caption{(Color online) Helmholtz free energy ${\cal A}$ vs. polarization $P$ and 
surface charge $\sigma$ for a
$t = 3.2$~nm PbTiO$_3$ film coherently strained to SrTiO$_3$ at 644~K, 
with $\lambda/t = 10^{-4}$.
Color scale gives values of ${\cal A}$.
To emphasize the equilibrium solutions, only the region near $P = -\sigma$ is plotted
since ${\cal A}$ is very large outside this region.}
%(chem\_ferroN2\_6.pdf)}
\label{F6aa}
\vspace{-0.1in}
\end{figure}

\subsection{Phase diagram for controlled $\sigma$}

\begin{figure}
\centering
\includegraphics[width=3.0in]{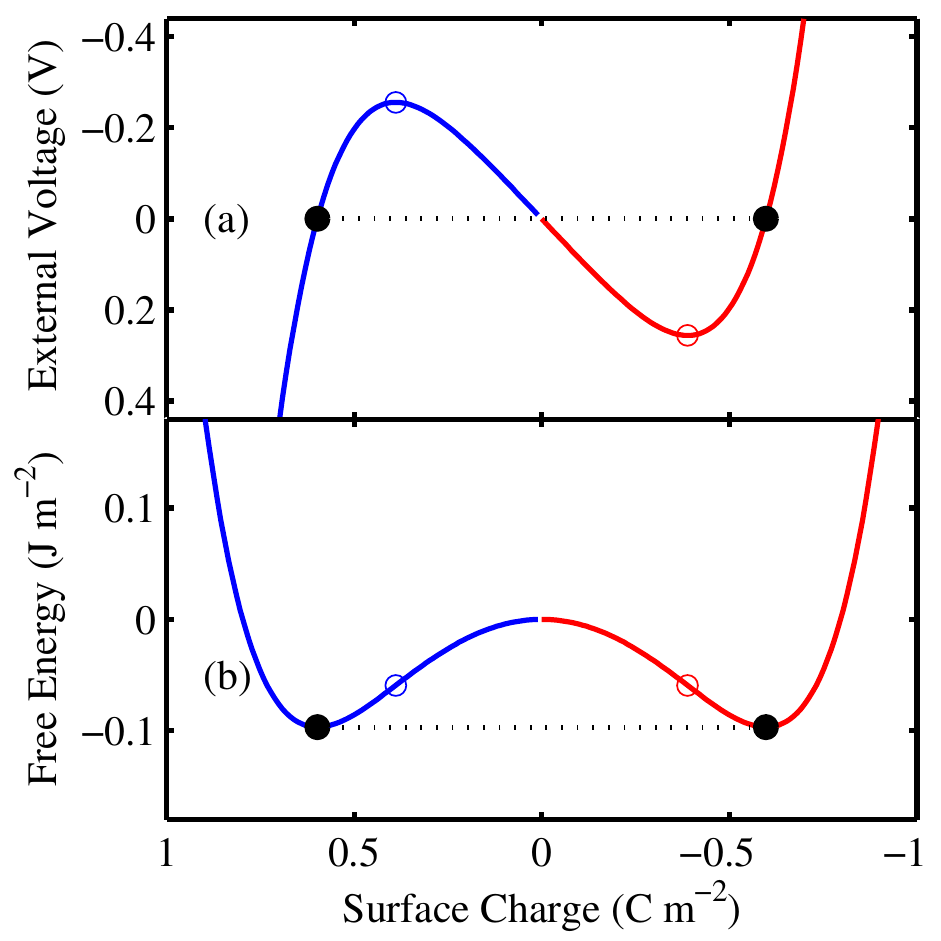} 
\caption{(Color online) Equilibrium solutions for (a) external voltage $V_{ex}$ and 
(b) Helmholtz free energy ${\cal A}$ as a function of surface charge $\sigma$
corresponding to Fig.~\ref{F6aa}. 
Red (blue) curves are positive (negative) polarization; 
all values are stable with respect to $P$ variations when $\sigma$ is spatially uniform.
Closed (open) circles show the equilibrium transition (instability) points
when $\sigma$ can be nonuniform.}
%(chem\_ferroN2\_4.pdf)}
\label{F6a}
\vspace{-0.1in}
\end{figure}

An alternative to controlling the voltage $V_{ex}$ across the electrodes
is to control the current flow to the electrodes, and hence the free charge $\sigma$.
The equilibrium polarization at fixed $\sigma$ is determined by minimizing the
Helmholtz free energy for the system
given by  Eq.~(\ref{eq5a}).
This is plotted as a function of $P$ and $\sigma$ in Fig.~\ref{F6aa}.
Setting to zero the derivative of $\cal A$ at constant $\sigma$
gives the equation of state
\begin{equation}
0 = \frac{1}{t} \left . \frac{\partial \cal A}{\partial P} \right | _ {\sigma}
= f^\prime (P) + \frac{\sigma + P}{\epsilon_0}.
\label{eq5i}
\end{equation}
This agrees with the constitutive relation, Eq.~(\ref{eq5}),
verifying that the correct equilibrium states are predicted by this free energy
for fixed $\sigma$.
Since $C/T_C^{\circ}$ is very large for typical ferroelectrics,
there is a only one minimum at each value of $\sigma$.
The equilibrium values of $\cal A$ and $V_{ex}$ are plotted versus $\sigma$
in Fig.~\ref{F6a}.
To a good approximation, this equilibrium solution
corresponds to $P_{eq} \approx -\sigma$, and
\begin{eqnarray}
{\cal A}_{eq} &\approx& t f(\sigma) + \frac{ \lambda \sigma^2}{\epsilon_0}, \nonumber \\
V_{ex}^{eq} &\approx& t f^{\prime}(\sigma) + 2\lambda \frac{ \sigma}{\epsilon_0}.
\label{eq5ii}
\end{eqnarray}
If these approximate expressions were plotted with the exact results in Fig.~\ref{F6a},
the curves would be indistinguishable.

% figure 6a: Vex and Free energy as a function of $\sigma$

The stability of this equilibrium solution with respect to
variations in $P$ can be evaluated
from the sign of the second derivative of ${\cal A}$ with respect to $P$,
\begin{equation}
\frac{1}{t} \left . \frac{\partial^2 \cal A}{\partial P^2} \right | _ {\sigma}
= f^{\prime\prime} (P) + \frac{1}{\epsilon_0}.
\end{equation}
Because $C/T_C$ is very large,
the first term is negligible, and the second derivative is always positive.
Unlike the fixed $V_{ex}$ case, the equilibrium solution $P_{eq} \approx -\sigma$
is never unstable
with respect to fluctuations in $P$,
and there is no phase transition.
This point is often not recognized --
{\it if the surface charge is controlled and kept uniform,
any value of $P$ in the film may be stably formed.}

While at fixed surface charge density $\sigma$ the equilibrium solution is always stable 
with respect to polarization variations,
instability can occur with respect to spatial nonuniformity in $\sigma$.
The free energy ${\cal A}_{eq}$ of Eq.~(\ref{eq5ii}) is a double well,
as shown in Fig.~\ref{F6a}(b).
The minima occur at values $\sigma_{min}$ given by solutions of
\begin{equation}
0 = \frac{1}{t} \frac{d {\cal A}_{eq}}{d \sigma} 
\approx f^{\prime}(\sigma) + \frac{2\lambda \sigma}{t \epsilon_0}.
\end{equation}
This gives $V_{ex}^{eq}(\sigma_{min}) = 0$.
If the controlled parameter is the net charge density on the electrode ${\bar \sigma}$,
then for $|{\bar \sigma}| < |\sigma_{min}|$
the system can lower its free energy by forming
a two-phase mixture of domains of opposite polarity.
The extent of this two-phase region is shown by the black
dotted lines in Fig.~\ref{F6a}.
The local surface charge $\sigma = \pm |\sigma_{min}|$
will have opposite sign for oppositely polarized domains.
At equilibrium, the fraction of positive domains $x_{pos}$ will be
$x_{pos} = (1 - {\bar \sigma} / |\sigma_{min}|)/2$.
The equilibrium value of $V_{ex}$ is zero in this polydomain region
and the free energy densities of the oppositely polarized domains are equal.

Here we assume that the in-plane size of the domains 
is sufficiently large compared with the film thickness
that we can neglect the excess free energy of the domain walls
and the in-plane components and nonuniformity of the polarization and field 
near the domain walls.
When there is incomplete neutralization of the depolarizing field by compensating charge
(e.g. when $\lambda$ is not zero),
the free energy can in some cases be reduced by the formation of 
equilibrium 180$^{\circ}$ stripe domains\cite{02_StriefferPRL_89_067601,GBS06JAP,BRAT08JCTN}
with an in-plane size similar to or less than the film thickness.
For such fine-scale domain structures, the domain wall energies
and field variations are not negligible.
For simplicity, we do not consider these cases here.

The  surface charge density is a conserved order parameter,
so instability with respect to spatial nonuniformity
occurs for magnitudes of $\sigma$ smaller than the spinodal values
given by
\begin{equation}
0 = \frac{1}{t} \frac{d^2 {\cal A}_{eq}}{d \sigma^2} \approx f^{\prime\prime}(\sigma) 
+ \frac{2\lambda}{t \epsilon_0}.
\label{eq5gg}
\end{equation}
Since $\lambda << t$, this expression gives the same result
as the instability with respect to uniform $P$ variations at constant $V_{ex}$,
Eq.~(\ref{eq5g}).

\begin{figure}
\centering
\includegraphics[width=3.0in]{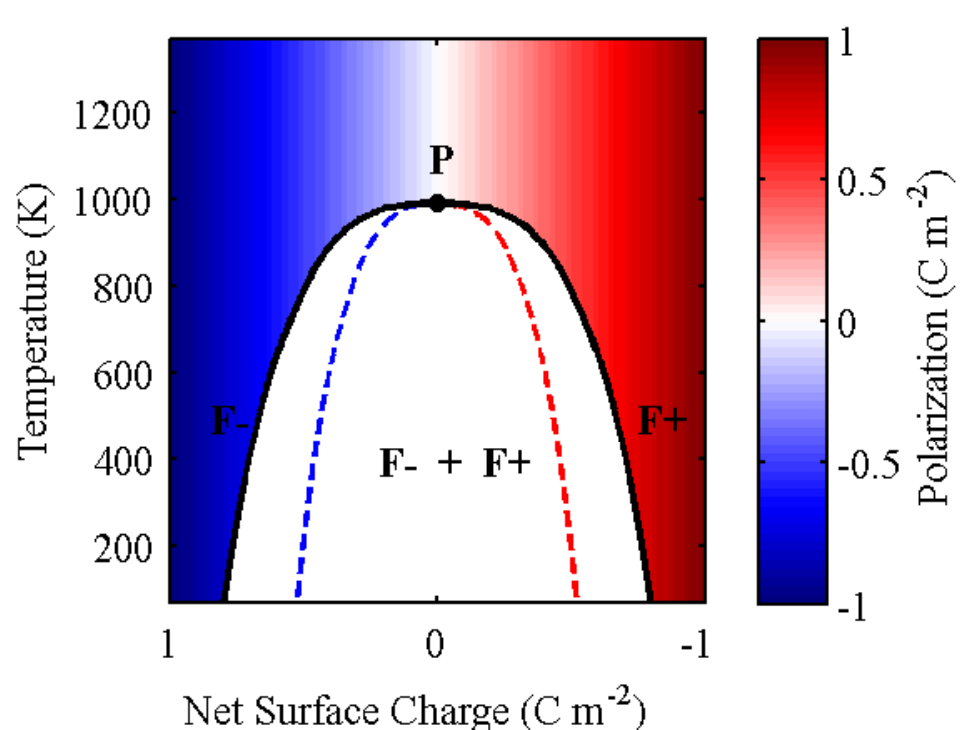} 
\caption{(Color online) Equilibrium phase diagram 
as a function of net surface charge ${\bar \sigma}$ 
and $T$ when $\sigma$ can be nonuniform, 
for PbTiO$_3$ coherently strained to SrTiO$_3$ 
with $t = 3.2$~nm and $\lambda/t = 10^{-4}$.
Color scale gives polarization in single-phase region.
Solid black line is phase boundary between 
positive and negative polar ferroelectric (F+ and F-) phases
and a two-phase field,
which terminates at $T_C$ (filled circle). 
Dashed red and blue curves are metastability limits 
of the F+ and F- phases, respectively.}
%(chem\_ferroNX6\_4.pdf)}
\label{F6d}
\vspace{-0.1in}
\end{figure}

\begin{figure}
\centering
\includegraphics[width=3.0in]{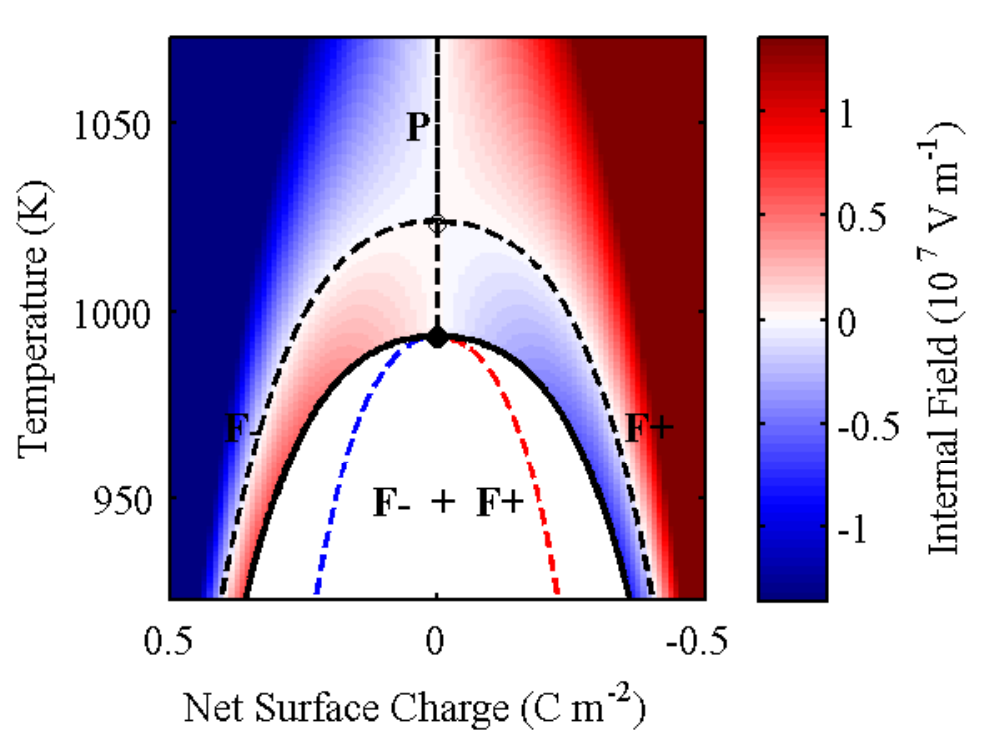} 
\caption{(Color online) Internal field 
in the region near the Curie point, corresponding to Fig.~\ref{F6d}.
Color scale gives electric field in single-phase region.
Dashed black curves show conditions for zero field.}
%(chem\_ferroNX7\_6.pdf)}
\label{F6e}
\vspace{-0.1in}
\end{figure}

Figure~\ref{F6d} shows the equilibrium polarization phase diagram 
as a function of ${\bar \sigma}$ and $T$,
while Fig.~\ref{F6e} shows the internal field in the vicinity of $T_C$.
These exhibit a two-phase field between the single-phase positive and negative
polar phases (F+ and F-).
The dashed red and blue spinodal curves are the metastability limits of the 
F+ and F- phases, respectively.
The polarization behavior is especially simple,
since we have
$P \approx -\sigma$
in the single-phase regions,
independent of $T$.
The suppression of $T_C$ is the same as in the controlled $V_{ex}$ case,
and the equilibrium and instability curves correspond exactly.
However, transformations driven by controlling either $V_{ex}$ or ${\bar \sigma}$
follow different paths.
If $V_{ex}$ is kept constant, the parent phase will remain metastable and will be
entirely consumed by the stable phase.
If ${\bar \sigma}$ is kept constant, $V_{ex}$ will decrease to zero as the
fraction of inverted domains grows,
reaching an equilibrium two-phase state.
The controlled potential and controlled charge phase diagrams
for ferroelectrics, Figs.~\ref{F2e} and~\ref{F6d},
are directly analogous to controlled chemical potential and controlled composition
phase diagrams for an alloy or fluid exhibiting phase separation.\cite{CAHN59JCP}
In particular,
the instability in ferroelectrics is a spinodal boundary,
and the continuous transition that occurs at fixed $V_{ex}$ in the unstable region
is equivalent to spinodal decomposition of an alloy held at constant chemical potential.
In this case, unlike the usual fixed average composition constraint for an alloy,
the continuous transition will result in a single-phase (monodomain) final state.
Spinodal transitions from monodomain to polydomain states in ferroelectrics 
at fixed ${\bar \sigma}$ have recently been modeled.\cite{ARTEMEV10AM}

% figure 6d: Polarization phase diagram as a function of $\sigma$ and T

% figure 6e: Field phase diagram as a function of $\sigma$ and T

While the conclusion that 
the $P \approx -\sigma$ solution is always stable
with respect to $P$ fluctuations
for any constant $\sigma$
may seem practically irrelevant 
for the electronic compensation case 
where the system is unstable with respect to $\sigma$ nonuniformity,
in the case of ionic compensation
this conclusion can be important.
As we shall see,
for ionic compensation the system can be stable 
against $\sigma$ nonuniformity, 
and the phase transition to a polar state can be completely suppressed
for a range of applied chemical potential.

\section{Ferroelectric film with ionic surface compensation}

Now we consider a ferroelectric film without a top electrode, 
but with its surface exposed to a chemical environment 
that can supply free charge from ionic species.
The amount of free charge supplied will depend on 
the chemical composition of the environment
and the external voltage $V_{ex}$ that it sees on the surface.
We will use the same 
electrostatic boundary condition, Eq.~(\ref{eq3}),
constitutive relation, Eq.~(\ref{eq5}),
and free energy, Eq.~(\ref{eq5a}),
employed above for the electronic compensation cases,
treating the ions as residing in a plane at a distance $\lambda$ 
above the surface.
Rather than solving for the polarization for a given value of $V_{ex}$
or $\sigma$,
we wish to obtain the equilibrium  polarization for a given 
composition of the environment.

To obtain the relationship between $\sigma$ and $V_{ex}$ 
due to this chemical equilibrium,
we develop an expression based on those for adsorption of ions
used in electrochemical systems.\cite{96_Schmickler}
We treat the external chemical environment as an electrolyte
that is in contact with both the surface of the film
and the bottom electrode
(e.g. via pinholes in the film away from the region of interest).
In order for surface ions from the chemical environment to produce an electric field across the sample,
as observed in experiments,\cite{09_WangPRL_102_047601,10_Kim_APL96_202902,10_HighlandPRL}
the electrons involved in creating the surface ions
must have such a path to reach the bottom electrode.

We can write a generalized surface redox reaction 
between oxygen in the environment
and a particular surface ion $i$,
\begin{equation}
{\rm Ion Site} + \frac{1}{n_i} {\rm O}_2 \leftrightarrow z_i e^- + {\rm Ion}^{z_i},
\label{eq15a}
\end{equation}
where $n_i$ is the number of surface ions created per oxygen molecule,
and $z_i$ is the charge on the surface ion.
In this formalism, $n_i$ and $z_i$ change sign depending upon whether
positively or negatively charged surface species are involved.
For example, if the surface ion is a
doubly-negatively-charged single-atom adsorbed oxygen, ${\rm O}_{ad}^{2-}$,
so that $n_i = 2$ and $z_i= -2$,
the redox reaction is
\begin{equation}
{\rm V}_{ad} + \frac{1}{2} {\rm O}_2 + 2 e^- \leftrightarrow {\rm O}_{ad}^{2-},
\label{eq16}
\end{equation}
while if the surface ion is a
doubly-positively-charged single-atom missing surface oxygen, ${\rm V}_O^{2+}$,
so that $n_i = -2$ and $z_i = 2$,
the redox reaction is
\begin{equation}
{\rm O}_O \leftrightarrow \frac{1}{2} {\rm O}_2 + 2 e^- + {\rm V}_O^{2+}.
\label{eq17}
\end{equation}
In these reactions ${\rm V}_{ad}$ represents a vacant oxygen ion 
adsorption site on top of the film
and ${\rm O}_O$ represents an occupied oxygen site 
in the outermost layer of the film.
We include these sites in the equilibrium so that the concentration of ions
saturates when all sites in the relevant surface layer are filled.
The concentrations of surface ions,
$\theta_i \equiv [{\rm Ion}]$, 
are defined so that their saturation levels are $\theta_i = 1$ 
and the concentrations of the surface sites are 
$[{\rm Ion Site}] = 1 - \theta_i$. 

One can write mass-action equilibria for these redox reactions, 
taking into account the external voltage difference 
between the bottom electrode and the surface 
(since the electrons are assumed to reside at the bottom electrode,
while the ions reside at the surface).
These are given by
\begin{equation}
\frac{\theta_i}{1 - \theta_i} = p_{O_2}^{1/n_i} \exp \left ( \frac{-\Delta G_i^{\circ}  
- z_i e V_{ex}}{kT} \right ),
\label{eq18}
\end{equation}
where  $\Delta G_i^{\circ}$ is the standard free energy of formation of the
surface ion at $p_{O_2} = 1$~bar and $V_{ex} = 0$, 
and $e$ is the magnitude of the electron charge.
This expression is analogous to the Langmuir adsorption isotherm
used in interfacial electrochemistry\cite{96_Schmickler}
for adsorption of neutral species onto a conducting electrode
exposed to ions in a solution.
Here, we consider adsorption of ions onto a polar surface
exposed to neutral species in a chemical environment.
Thus our $V_{ex}$ is the potential of the adsorbed ions relative to the electrons,
rather than the potential of the electrons relative to the ions in solution
as in the typical electrochemical case.

The standard free energies can depend not only on temperature
but also on the polarization of the film, 
since the surface structure changes with $P$.
For simplicity
we assume that they can be all described by the same parameter 
$\lambda^{\prime}$ using
\begin{eqnarray}
\Delta G_i^{\circ} &\equiv& \Delta G_i^{\circ\circ}(T) 
+ (z_i e \lambda^{\prime}/\epsilon_0) P.
\label{eq25}
\end{eqnarray}
If $\lambda^{\prime}$ is positive,
then a more positively polarized film tends to stabilize negative surface ions,
and vice versa.
Note that the effect of $\lambda^{\prime}$ is in addition to the electrostatic energy
already included through the $z_i e V_{ex}$ term in Eq.~(\ref{eq18}).
The density of free charge on the surface is the sum of those 
from the various surface ions,
giving
\begin{equation}
\sigma = \sum_i \frac{z_i e \theta_i}{A_i},
\label{eq20}
\end{equation}
where the $A_i^{-1}$ are the saturation densities of the surface ions.

\begin{table}
\caption{Values of ionic surface compensation coefficients used in displayed plots.}
\label{tab2}
\begin{ruledtabular}
\begin{tabular}{cccccc}

$\Delta G_\op^{\circ\circ}$ &1.00 &(eV) 
&$\Delta G_\om^{\circ\circ}$ &0.00 &(eV) \\ 

$n_\op$ &$-2$ &
&$n_\om$ &$2$ & \\ 

$z_\op$ &$2$ & 
&$z_\om$ &$-2$ & \\ 

$A_\op$ &$1.6 \times 10^{-19}$ & (m$^2$) 
&$A_\om$ &$1.6 \times 10^{-19}$ & (m$^2$) \\ 

$\lambda$ & 0 &(m) & $\lambda^{\prime}$ &0 &(m)\\ 

\end{tabular}
\end{ruledtabular}
\end{table}

Using Eqs.~(\ref{eq18})-(\ref{eq20}) and Eq.~(\ref{eq3}),
the surface ion concentrations $\theta_i$
can be calculated for given $V_{ex}$ and $p_{O_2}$. 
Parameter values for a system with one positive surface ion $i = \op$ 
and one negative surface ion $i = \om$
are given in Table~\ref{tab2};
here $\Delta G_\op^{\circ\circ}$ and $\Delta G_\om^{\circ\circ}$
are taken to be independent of temperature.
We have assumed that the saturation densities of the surface ions $A_i^{-1}$
are both one per PbTiO$_3$ unit cell area.
For divalent surface ions, 
this saturation density would provide more than twice the charge density 
needed to fully compensate the typical polarization of PbTiO$_3$.

% figure 4: surface defect concentrations
\begin{figure}
\centering
\includegraphics[width=3.0in]{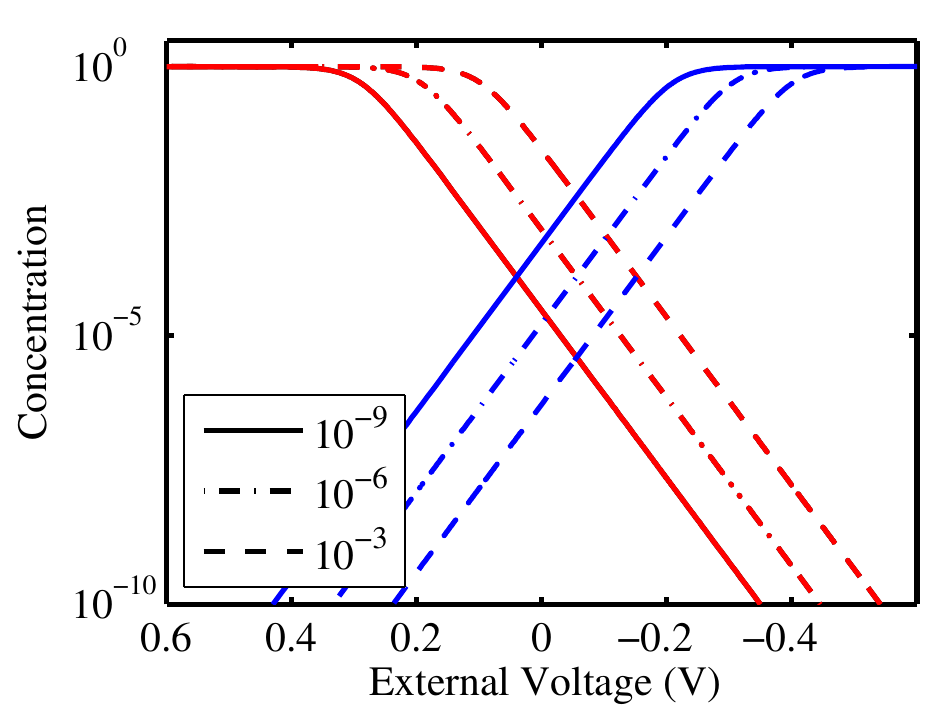} 
\caption{(Color online) Concentrations of positive $\theta_\op$  (red curves) 
and negative $\theta_\om$ (blue curves) surface ions
as a function of $V_{ex}$,
calculated for the values of $ p_{O_2}$ (bar) given in the legend
at $T = 644$~K, $t = 3.2$~nm. 
Parameter values used are
given in Table~\ref{tab2}.}
% (chem\_ferroNB6\_2.pdf)}
\label{F4}
\vspace{-0.1in}
\end{figure}

% figure 5: surface charge
\begin{figure}
\centering
\includegraphics[width=3.0in]{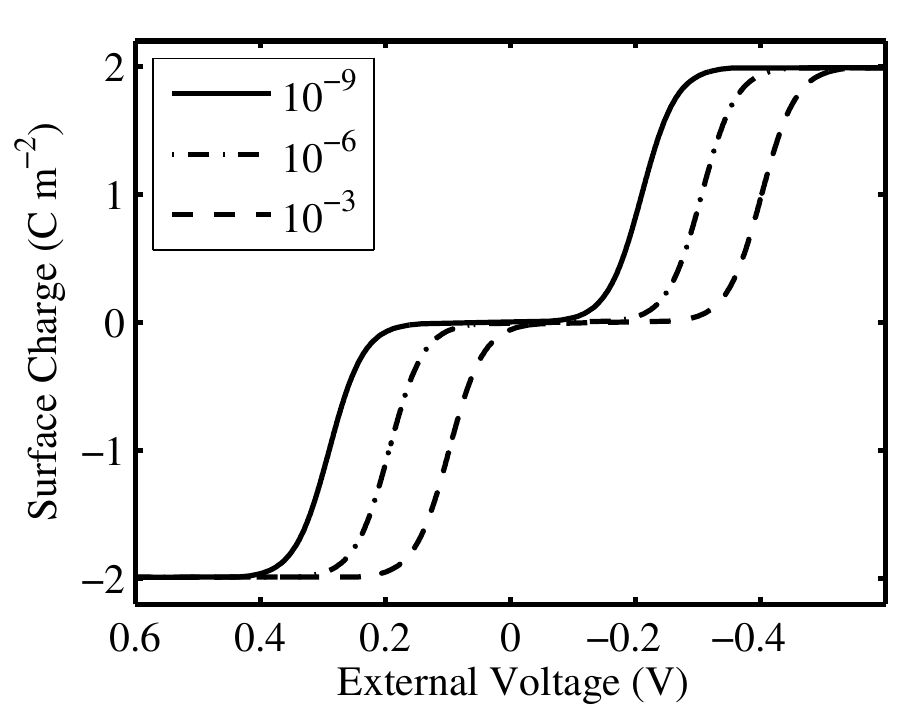} 
\caption{Surface charge density $\sigma$
as a function of $V_{ex}$ at three values of $p_{O_2}$,
corresponding to the concentrations shown in Fig.~\ref{F4}. 
Parameter values used are
given in Table~\ref{tab2}.}
% (chem\_ferroNB6\_1.pdf)}
\label{F5}
\vspace{-0.1in}
\end{figure}

Figure~\ref{F4} shows $\theta_i$
as a function of $V_{ex}$
for several different values of $p_{O_2}$.
Changing $p_{O_2}$ shifts the external voltage scale by
$[kT/(z_i n_i e)] \Delta \ln p_{O_2}$
for each ion.
Figure~\ref{F5} shows the corresponding surface charge density $\sigma$.
Three plateaus occur --
two at extreme values of $V_{ex}$, 
where one or the other of the ionic surface concentrations $\theta_i$ 
saturates at unity,
and a third near zero surface charge over the range of $V_{ex}$
for which both $\theta_i$ are small compared to unity.
From the shape of the charge vs. voltage curves in Fig.~\ref{F5},
one can see that this fixed $p_{O_2}$ boundary condition
has regions that correspond to fixed $\sigma$
separated by regions that correspond approximately to fixed $V_{ex}$.
Thus fixed $p_{O_2}$ does not correspond to either fixed $\sigma$
or fixed $V_{ex}$.
As we shall see,
this strongly affects the equilibrium phase diagram.

The values of $p_{O_2}$ and $V_{ex}$ that give $\sigma = 0$ 
can be obtained by solving Eqs.~(\ref{eq18})-(\ref{eq20})
for $\sigma = P = 0$, $\theta_i << 1$, to give
\begin{equation}
\ln \left ( \frac{p_{O_2}}{p^{\circ\circ}_{O_2}} \right ) 
= \frac{n_\om n_\op (z_\op - z_\om)}{n_\om - n_\op} \frac{eV_{ex}}{kT},
\label{eq31c}
\end{equation}
where $p^{\circ\circ}_{O_2}$ is the temperature-dependent
oxygen partial pressure that gives 
$\sigma = P = 0$ at $V_{ex} = 0$,
\begin{equation}
\ln p^{\circ\circ}_{O_2} \equiv \frac{-n_\om n_\op}{n_\om - n_\op}  
\left [ \frac{\Delta G_\om^{\circ\circ} - \Delta G_\op^{\circ\circ}}{kT} 
+ \ln \left ( \frac{z_\op A_\om}{-z_\om A_\op} \right ) \right ]. 
\label{eq31d}
\end{equation}
As shown below, this value of $p_{O_2}$ 
marks the transition between oppositely polarized films
on the phase diagram.

\subsection{Equilibrium solutions at controlled oxygen partial pressure}

\begin{figure}
\centering
\includegraphics[width=3.0in]{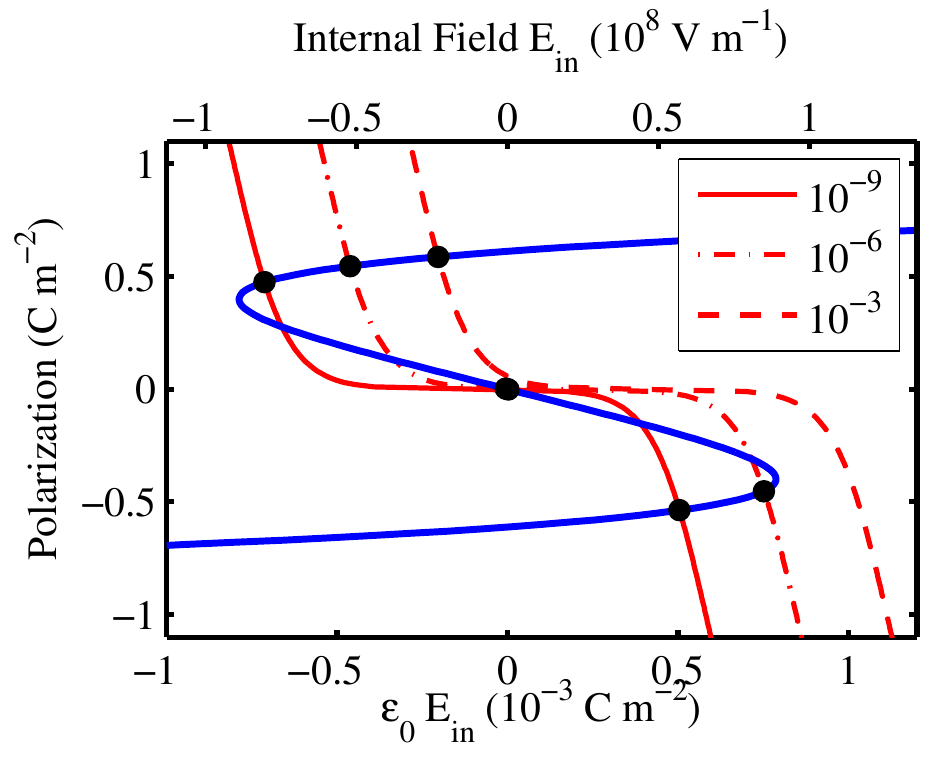} 
\caption{(Color online) Polarization vs. internal field relationships arising 
from the constitutive relation for the ferroelectric film (blue ``S'' shaped curve)
and
from the chemical boundary
condition (red curves with plateau) for $p_{O_2}$ values shown in legend (bar).
Marked intersections correspond to stable or metastable equilibrium solutions.
Parameter values used are
given in Tables~\ref{tab1} and~\ref{tab2}, with
$T = 644$~K,
$t = 3.2$~nm.}
% (chem\_ferroNB6\_3.pdf)}
\label{F7}
\vspace{-0.1in}
\end{figure}

Equilibrium solutions can be calculated by obtaining a relationship between
$E_{in}$ and $P$ due to the fixed $p_{O_2}$ chemical boundary condition,
and solving it simultaneously with the constitutive relation for the ferroelectric,
as we did for the electronic case in Fig.~\ref{F2}.
The solution for $\sigma$ as a function of $V_{ex}$ shown in Fig.~\ref{F5}
along with Eqs.~(\ref{eq3}) and (\ref{eq4})
gives a relation between $E_{in}$ and $P$ for a given $p_{O_2}$,
which can be solved simultaneously with Eq.~(\ref{eq5})
to obtain the overall equilibrium.
This is illustrated in Fig.~\ref{F7},
where the chemical boundary condition
for three values of $p_{O_2}$ is shown.
Changing $p_{O_2}$ shifts the boundary condition curve along the $E_{in}$ axis.
The boundary condition curve is centered on $P = 0$, $E_{in} = 0$
when $p_{O_2}$ is equal to $p^{\circ\circ}_{O_2}$ of Eq.~(\ref{eq31c}).

% figure 7: P vs E

An approximate expression for the chemical boundary condition can be obtained 
by making some simplifying assumptions.
The dependence of $\sigma$ on $V_{ex}$ obtained from the depolarizing field, 
Eq.~(\ref{eq3}), 
is negligible compared with that obtained from the chemical equilibria,
Eq.~(\ref{eq20}), shown in Fig.~\ref{F5}.
One can approximate Eq.~(\ref{eq3}) as $\sigma \approx -P$.
In addition,
we can neglect one of the ion concentrations $\theta_\op$ or $\theta_\om$ 
relative to the other,
depending on the sign of the film polarization.
This leads to the limiting expressions
\begin{equation}
V_{ex} \approx - \frac{kT}{z_i e} \ln \frac{-A_i P / z_i e}{1 + A_i P / z_i e}  
- \frac{ \Delta G_i^{\circ\circ}}{z_i e}  - \frac{\lambda^{\prime}}{\epsilon_0} P 
+  \frac{kT \ln p_{O_2}}{z_i n_i e} ,
\label{eq21}
\end{equation}
for $i = \om$ or $\op$ (positive or negative film polarization, respectively).
For the parameters used in Fig.~\ref{F5}, e.g. far below $T_C$,
the approximation (\ref{eq21}) is very close to the exact solution
obtained numerically. 
Substituting the approximation (\ref{eq21}) into Eq.~(\ref{eq4}),
one obtains relationships between field and polarization given by
\begin{equation}
\epsilon_0 E_{in} \approx \frac{\epsilon_0 (\frac{kT}{z_i e} \ln \frac{-A_i P / z_i e}{1 + A_i P / z_i e} 
+ \frac{\Delta G_i^{\circ\circ}}{z_i e} - \frac{kT \ln p_{O_2}}{z_i n_i e}) 
- 2\lambda^{\dagger} P}{2\lambda + t},
\label{eq30}
\end{equation}
for $i = \om$ or $\op$.
Here we have introduced a new parameter, $\lambda^{\dagger}$, defined by
\begin{equation}
\lambda^{\dagger} \equiv \lambda - \lambda^{\prime} /2 .
\end{equation}
If $\lambda^{\prime}$ is positive then $\lambda^{\dagger}$
is smaller than $\lambda$,
and in particular $\lambda^{\dagger}$ can be negative.

\subsection{Stability of equilibrium solutions}

Because of the plateau in the $P(E_{in})$ shape of the chemical boundary condition
shown in Fig.~\ref{F7},
there can be as many as five equilibrium solutions given by the intersections.
A total free energy function that is minimized at equilibrium can be used
to determine which solutions are stable, metastable, and unstable.
The Gibbs free energy consistent with the above treatment of ionic surface compensation is
\begin{eqnarray}
{\cal G} &=&  t \left [ f(P) +\frac{ (\sigma + P)^2}{2\epsilon_0} \right ] 
+ \frac{\lambda^{\dagger} \sigma^2}{\epsilon_0} \nonumber \\
&+& \sum_{i=\om,\op} \frac{kT}{A_i} \left [  \frac{\theta_i \Delta G_i^{\circ\circ}}{kT} 
 - \frac{\theta_i \ln p_{O_2}}{n_i} \right . \nonumber \\
&+& \biggl. \theta_i \ln \theta_i + (1-\theta_i) \ln (1-\theta_i) \biggr].
\label{eq20a}
\end{eqnarray}
Minimizing this ${\cal G}$ with respect to $P$ at
constant $p_{O_2}$, $\theta_\om$, and $\theta_\op$ 
(and therefore constant $\sigma$) gives
\begin{equation}
0 = \frac{1}{t} \left . \frac{\partial {\cal G}}{\partial P} \right | _ {\theta_\op, \theta_\om, p_{O_2}}
= f^\prime (P) + \frac{\sigma + P}{\epsilon_0}.
\end{equation}
This agrees with the constitutive relation, Eq.~(\ref{eq5}),
like the case for electronic compensation,
Eq.~(\ref{eq5i}).
As in that case, 
because of the large value of $C/T_C$ for PbTiO$_3$,
the equilibrium polarization is given to a good approximation by $P \approx -\sigma$.
The free energy expression then becomes
\begin{eqnarray}
{\cal G} &\approx&  t  f(\sigma)
+ \frac{\lambda^{\dagger} \sigma^2}{\epsilon_0} \nonumber \\
&+& \sum_{i=\om,\op} \frac{kT}{A_i} \left [  \frac{\theta_i \Delta G_i^{\circ\circ}}{kT} 
- \frac{\theta_i \ln p_{O_2}}{n_i} \right . \nonumber \\
&+& \biggl. \theta_i \ln \theta_i + (1-\theta_i) \ln (1-\theta_i) \biggr] .
\label{eq20aa}
\end{eqnarray}
When the derivatives of this free energy with respect to $\theta_\op$ and $\theta_\om$
at fixed $p_{O_2}$ are set to zero,
this yields the equilibrium relations given above in Eqs.~(\ref{eq18}-\ref{eq20}),
where $V_{ex}$ of Eq.~(\ref{eq3})
is now given by
$V_{ex} \approx 2\lambda \sigma / \epsilon_0 + t f^{\prime}(\sigma)$.

The global minimum of the free energy ${\cal G}$ of Eq.~(\ref{eq20aa})
with respect to $\theta_\om$ and $\theta_\op$ 
typically occurs either at $\theta_\om \approx 0$ or $\theta_\op \approx 0$.
The generality of this result can be evaluated by re-expressing
the $\theta_\om$ and $\theta_\op$ terms in the free energy
Eq.~(\ref{eq20aa}) using new variables $\sigma$ and 
$\delta \equiv z_\op e \theta_\op / A_\op - z_\om e \theta_\om / A_\om$.
Minimizing ${\cal G}$ with respect to $\delta$ at fixed $\sigma$ gives
\begin{eqnarray}
0 &=& 2e \left . \frac{\partial {\cal G}}{\partial \delta} \right | _ {\sigma, p_{O_2}}
= \left ( \frac{\Delta G_\op^{\circ\circ}}{z_\op} - \frac{\Delta G_\om^{\circ\circ}}{z_\om} \right ) 
\nonumber \\
&+& \left [ \frac{kT}{z_\op} \ln \left ( \frac{\theta_\op}{1-\theta_\op} \right ) 
-  \frac{kT}{z_\om} \ln \left ( \frac{\theta_\om}{1-\theta_\om} \right ) \right ] \nonumber \\
&+& \left ( \frac{z_\op n_\op - z_\om n_\om}{z_\op n_\op z_\om n_\om} \right ) kT \ln p_{O_2} .
\label{eq20aaa}
\end{eqnarray}
The first term is positive in cases such as the one we consider,
when there is a region of intermediate $p_{O_2}$ with low concentrations
of both positive and negative surface ions.
The third term is zero for typical values of the $z_i$ and $n_i$.
Thus the equilibrium condition requires that the second term be negative,
which occurs only when either $\theta_\om$ or $\theta_\op$ is very small.
Substituting this result into Eq.~(\ref{eq20aa}) gives
\begin{eqnarray}
\label{eq22}
&{\cal G}& \approx  t  f(\sigma)
+ \frac{\lambda^{\dagger} \sigma^2}{\epsilon_0}
+ \frac{\sigma}{z_i e} \left ( \Delta G_i^{\circ\circ}
- \frac{kT \ln p_{O_2}}{n_i} \right ) \\
&+& \frac{kT}{A_i} \left [ \frac{A_i \sigma}{z_i e} \ln \left ( \frac{A_i \sigma}{z_i e} \right )
+ \left ( 1- \frac{A_i \sigma}{z_i e} \right ) \ln \left ( 1 - \frac{A_i \sigma}{z_i e} \right ) \right ], 
\nonumber
\end{eqnarray}
where $i = \om$ for $\theta_\op \approx 0$, positive $P$, and negative $\sigma$, 
or $i = \op$ for $\theta_\om \approx 0$, negative $P$, and positive $\sigma$. 
The third term is proportional to $\sigma$,
like a field term,
but the constant of proportionality changes when $\sigma$ changes sign
and the ionic species at the surface change between positive and negative ions.
This change in slope of ${\cal G}(\sigma)$ at $\sigma = 0$
can produce a stable or metastable minimum.

% figure 10d: free energy vs sigma and pO2
\begin{figure}
\centering
\includegraphics[width=3.0in]{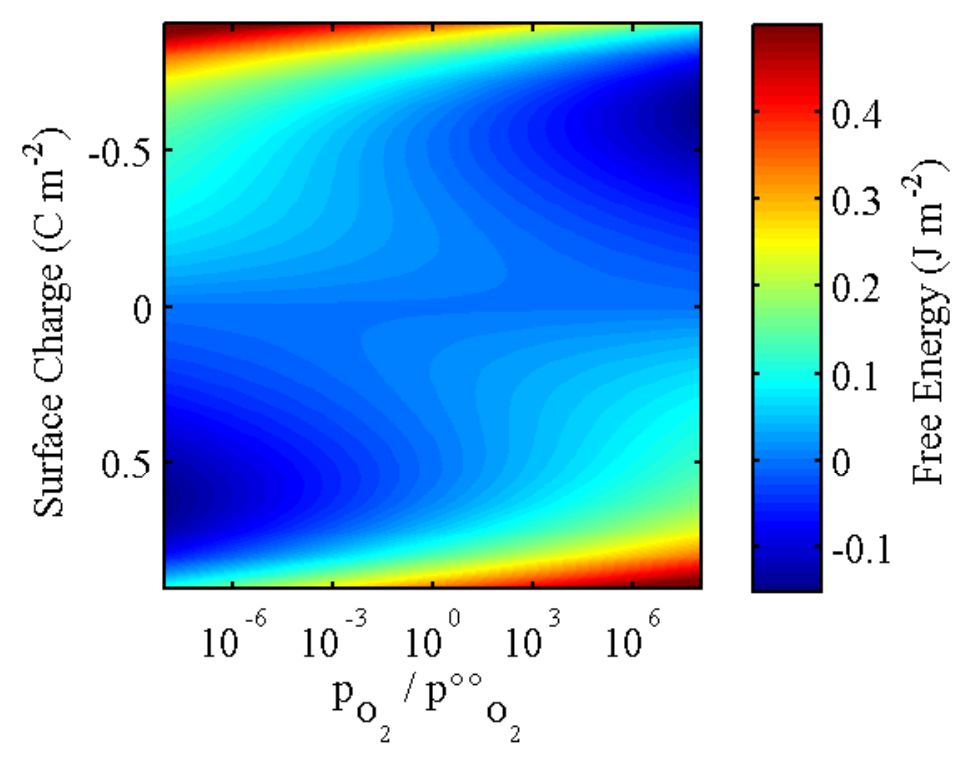} 
\caption{(Color online) Free energy as a function of $\sigma$
and $p_{O_2} / p^{\circ\circ}_{O_2}$.
Parameter values used are
given in Tables~\ref{tab1} and~\ref{tab2}, with
$T = 644$~K,
$t = 3.2$~nm.  
Color scale gives values of ${\cal G}$.}
% (chem\_ferroNB6\_6.pdf) }
\label{F10d}
\vspace{-0.1in}
\end{figure}

% figure 10e: free energy vs sigma and pO2
\begin{figure}
\centering
\includegraphics[width=3.0in]{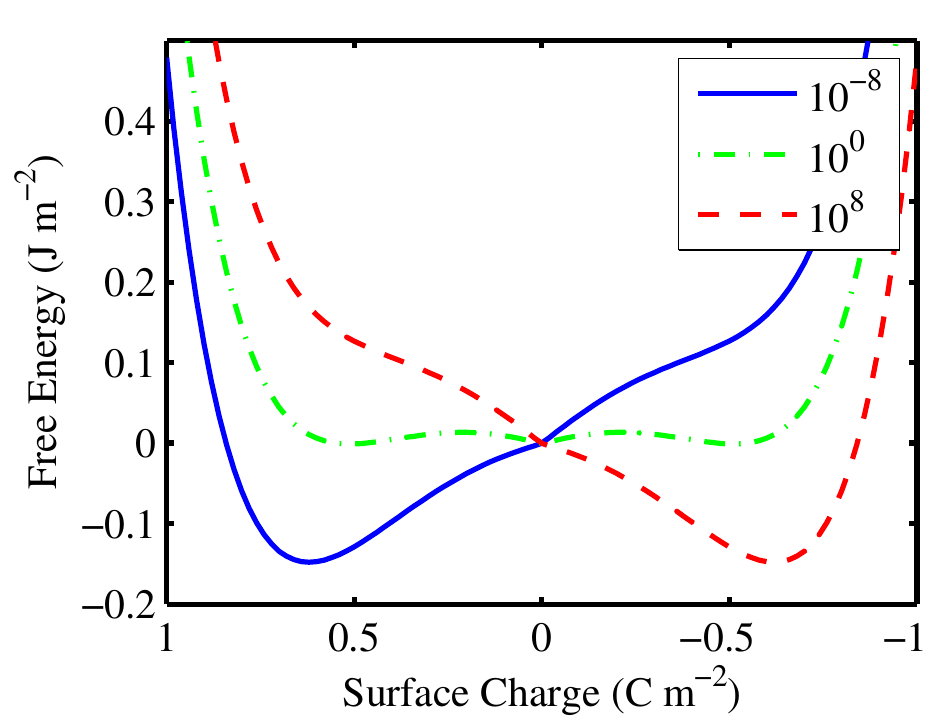} 
\caption{(Color online) Free energy as a function of $\sigma$
at three given values of $p_{O_2} / p^{\circ\circ}_{O_2}$.
Parameter values used are
given in Tables~\ref{tab1} and~\ref{tab2}, with
$T = 644$~K,
$t = 3.2$~nm.} 
% (chem\_ferroNB6\_7.pdf)}
\label{F10e}
\vspace{-0.1in}
\end{figure}

% figure 10f: free energy vs pO2
\begin{figure}
\centering
\includegraphics[width=3.0in]{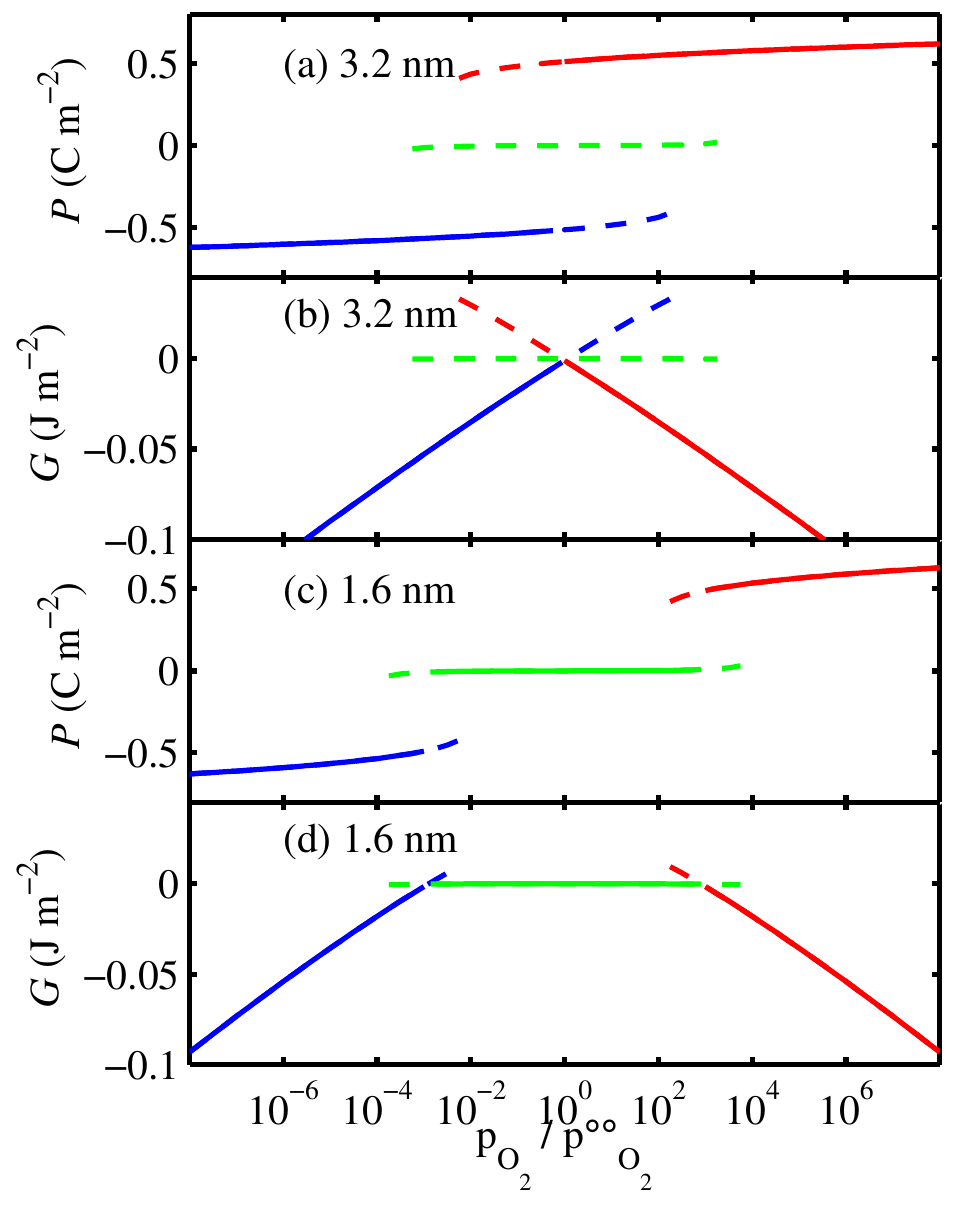} 
\caption{(Color online) Polarization $P$ and free energy ${\cal G}$ of the
(meta-)stable equilibrium solutions as a function of $p_{O_2} / p^{\circ\circ}_{O_2}$. 
Blue, green, and red curves are for negative, zero, and positive polarization solutions;
solid and dashed regions are stable and metastable,
respectively.
Parameter values used are
given in Tables~\ref{tab1} and~\ref{tab2}, with
$T = 644$~K,
and $t = 3.2$~nm for plots (a) and (b),
$t = 1.6$~nm for plots (c) and (d).}
% (chem\_ferroNB6\_4.pdf) }
\label{F10f}
\vspace{-0.1in}
\end{figure}

% figure 19a: T vs pO2 phase diagrams for 2.5, 5, and 10 nm films
\begin{figure*}
\centering
\includegraphics[width=7.0in]{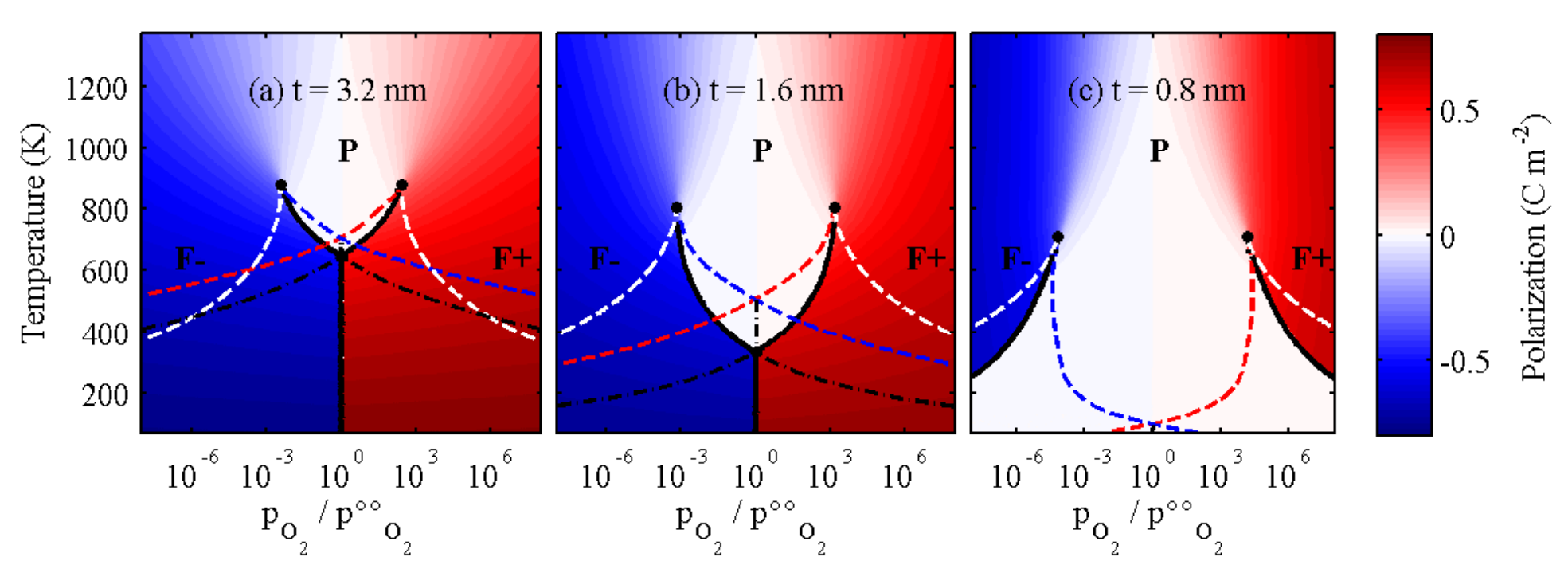} 
\caption{(Color online) Phase diagrams
as a function of $p_{O_2} / p^{\circ\circ}_{O_2}$ and $T$ for three thicknesses of PbTiO$_3$ 
coherently strained to SrTiO$_3$,
using parameters in Tables~\ref{tab1} and~\ref{tab2}.
Color scale gives polarization.
Solid and dash-dot black curves are stable and metastable phase boundaries,
respectively, between nonpolar paraelectric (P)
and positive and negative polar ferroelectric (F+ and F-) phases. 
Dashed red, blue, and white curves are metastability limits of the F+, F-, and P phases, 
respectively.}
%(chem\_ferroNTC3I\_19.pdf)}
\label{F19a}
\vspace{-0.1in}
\end{figure*}

The free energy of Eq.~(\ref{eq22})
is plotted versus $p_{O_2} / p^{\circ\circ}_{O_2}$ and $\sigma$ in
Figs.~\ref{F10d} and \ref{F10e}.
At intermediate values of $p_{O_2}$,
there are three (meta-)stable solutions corresponding to 
local minima in ${\cal G}(\sigma)$,
at positive, zero, and negative polarization.
These equilibrium solutions satisfy the equations of state
\begin{eqnarray}
0 &=& \frac{\partial {\cal G}}{\partial \sigma} \approx  t  f^{\prime}(\sigma)
+ \frac{2\lambda^{\dagger} \sigma}{\epsilon_0}
+ \frac{1}{z_i e} \left ( \Delta G_i^{\circ\circ}
- \frac{kT \ln p_{O_2}}{n_i} \right ) \nonumber \\
&+& \frac{kT}{z_i e} \left [ \ln \left ( \frac{A_i \sigma}{z_i e} \right )
- \ln \left ( 1 - \frac{A_i \sigma}{z_i e} \right ) \right ],
\label{eq22a}
\end{eqnarray}
and the limits of metastability of these solutions can be obtained from
\begin{equation}
0 = \frac{\partial^2 {\cal G}}{\partial \sigma^2} \approx  t  f^{\prime\prime}(\sigma)
+ \frac{2\lambda^{\dagger}}{\epsilon_0}
+ \frac{kT}{z_i e \sigma \left ( 1 - \frac{A_i \sigma}{z_i e} \right )},
\label{eq22b}
\end{equation}
where $i = \om$ or $\op$ as in Eq.~(\ref{eq22}).
Figure~\ref{F10f}(a,b) shows the polarizations and energies of these solutions
as a function of $p_{O_2} / p^{\circ\circ}_{O_2}$.
The energy of the solution at $P = 0$ is  zero,
while the energies depend on $p_{O_2} / p^{\circ\circ}_{O_2}$ for the other two solutions.
For the parameters used here, 
e.g. a 3.2~nm film thickness,
the energies of all three solutions are almost equal at 
$p_{O_2} / p^{\circ\circ}_{O_2} = 1$.
The $P = 0$ solution will be the stable (global minimum) solution
for thinner films at intermediate $p_{O_2}$.
Here this solution is stable against $\sigma$ nonuniformity,
unlike the electronic compensation case.
For example, Fig.~\ref{F10f}(c,d) shows the results
for a 1.6~nm thick film, with all other parameters the same.
In such thin films, where the central flat region of the boundary condition
$P(E_{in})$ curve spans a large range of $E_{in}$,
the positive and negative solutions do not overlap,
and the $P = 0$ solution is the only solution for the range of $p_{O_2}$
where both $\theta_\op$ and $\theta_\om$ are small.

For the polar phases, the last term of Eq.~(\ref{eq22b}) is typically small enough, 
except near $T_C$,
that this condition for the instability is very similar 
to those for the electronic compensation cases,
Eqs.~(\ref{eq5g}) and (\ref{eq5gg}).
Thus at the metastability limit of the polar phases, 
the internal field reaches the same intrinsic coercive field 
in the ionic compensation case as it does in the electronic compensation cases.

\subsection{Phase diagram for controlled $p_{O_2}$}

The effect of ionic surface compensation on the ferroelectric phase transition
can be explored by solving for the polarization and field as a function of temperature
as well as $p_{O_2}$ and film thickness.
As can be guessed from the fixed-temperature results shown above, 
the temperature dependences of $P$, $E_{in}$,
and the Curie point $T_C$ 
(i.e. the temperature of the equilibrium boundary between the
polar and nonpolar phases)
all vary with the $p_{O_2}$ of the environment.

Figure~\ref{F19a} shows equilibrium polarization phase diagrams as a function of 
$T$ and $p_{O_2}/p^{\circ\circ}_{O_2}$ for various film thicknesses.
In addition to the stable and metastable equilibrium phase boundaries,
the metastability limits of the polar and nonpolar phases are shown.
These phase diagrams are calculated using parameter values given in Table~\ref{tab2}.
The oxygen pressure scale has been normalized to $p^{\circ\circ}_{O_2}(T)$,
which produces symmetric diagrams
when $n_\om = - n_\op$, $z_\om = - z_\op$. 

The equilibrium phase diagrams as a function of $p_{O_2}$
for ionic compensation, Fig.~\ref{F19a},
differ qualitatively from the standard second-order ferroelectric phase diagrams
as a function of $V_{ex}$ or $\sigma$ for electronic compensation,
Figs.~\ref{F2e} and~\ref{F6d}.
The ionic phase diagrams show temperature ranges where
the nonpolar phase is stable at intermediate $p_{O_2}$
separating the positive and negative polar phases at high and low $p_{O_2}$,
respectively.
As film thickness becomes smaller, this ``wedge'' of nonpolar phase
extends to lower temperature, reaching 0~K for thicknesses
less than about 1~nm for the parameter values used here.
For films with smaller thickness, an inverted ferroelectric transition remains at extreme
values of $p_{O_2}$,
with the polar phase stable at temperatures above the phase boundary,
and the nonpolar phase stable below the boundary.
For thicker films,
there is a triple point where the first order transitions
between the positive and negative polar and nonpolar phases
meet at $p_{O_2} / p^{\circ\circ}_{O_2} = 1$,
while at extreme values of $p_{O_2}$
there is no phase transition as a function of $T$
between the polar and nonpolar phases,
similar to the case at nonzero $V_{ex}$ in Fig.~\ref{F2e}
for electronic compensation.
For all film thicknesses, the high temperature ends of the polar/nonpolar phase boundaries
terminate at two critical points.
In a range of temperature below these critical points,
the regions of (meta)stability of the positive and negative polar phases do not overlap.
Here, switching transitions between oppositely polarized states
at fixed $T$ driven by changing $p_{O_2}$
must occur through an intermediate nonpolar state.

The appearance of the nonpolar phase between the polar phases at lower temperature 
is directly related to the non-linear dependence of surface charge $\sigma$ 
on $\ln(p_{O_2})$ at fixed $V_{ex}$. 
This has a plateau at a value near $\sigma = 0$ for intermediate $p_{O_2}$ values, 
where the concentrations of both positive and negative surface ions are small. 
The low value of $\sigma$ in this region can be insufficient to stabilize either polar phase.

% figure 20: Curie and instab Ts vs film thickness
\begin{figure}
\centering
\includegraphics[width=3.0in]{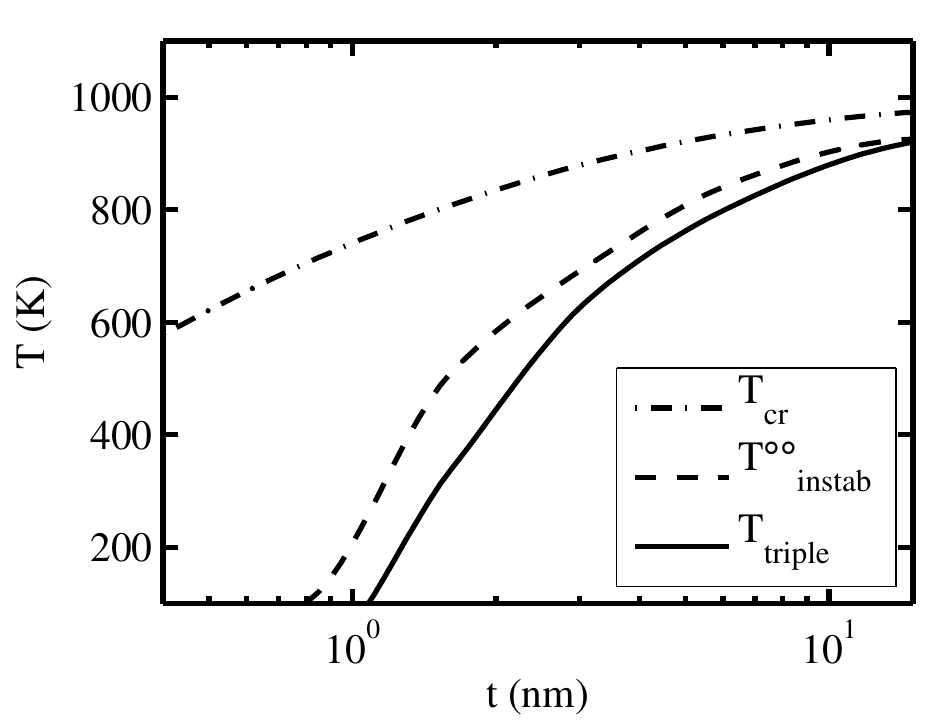} 
\caption{The temperatures of the critical points $T_{cr}$,
the polar phase instabilities at $p_{O_2} = p^{\circ\circ}_{O_2}$,
and the triple point,
as a function of film thickness.}
%(chem\_ferroNTC3L\_10.pdf)}
\label{F20}
\vspace{-0.1in}
\end{figure}

Figure~\ref{F20} shows the temperatures as a function of film thickness
of the critical points,
the polar phase instabilities at $p_{O_2}/p^{\circ\circ}_{O_2} = 1$,
and the triple point.
The triple point is the minimum equilibrium $T_C$. 
At temperatures between the triple point and the critical points, the nonpolar phase intervenes
between the polar phases at equilibrium.
An expression for the temperatures $T_{cr}$ of the critical points can be obtained 
by setting the second and third derivatives
of the free energy simultaneously to zero,
Eq.~(\ref{eq22b}) and
\begin{equation}
0 = \frac{\partial^3 {\cal G}}{\partial \sigma^3} \approx  t  f^{\prime\prime\prime}(\sigma)
- \frac{kT \left ( 1 - \frac{2A_i \sigma}{z_i e} \right )}{z_i e \sigma^2 \left ( 1 - \frac{A_i \sigma}{z_i e} \right )^2},
\label{eq22c}
\end{equation}
where $i = \om$ or $\op$ as in Eq.~(\ref{eq22}).
In the approximation that $\sigma$ is small at the critical point,
these reduce to
\begin{equation}
0 \approx \approx \left ( \alpha_3^*(T_{cr}) + \frac{\lambda^{\dagger}}{\epsilon_0 t} \right )^3
+ 2\alpha_{33}^*
 \left ( \frac{9kT_{cr}}{4 t | z_i | e} \right )^{2}.
\label{eq22d}
\end{equation}
We use the double $\approx \approx$ symbol to indicate
a rough approximation,
in this case because it becomes invalid at small $t$.
Nonetheless,
Eq.~(\ref{eq22d}) shows that the temperatures of the critical points for the ionic compensation case
are suppressed by an additional thickness-dependent term not present in the
electronic compensation case, Eq.~(\ref{eq5m}).
Even if the effective screening length is  zero, $\lambda^{\dagger} = 0$,
the $T_{cr}$ are changed by an amount
\begin{equation}
\Delta T_{cr} \equiv T_{cr} - T_C^{\circ} \approx \approx  -2\epsilon_0 C (2\alpha_{33}^*)^{1/3}
 \left ( \frac{9kT_{cr}}{4 t | z_i | e} \right )^{2/3} .
\end{equation}
Using the LGD parameters\cite{GBS06JAP} for PbTiO$_3$ coherently strained to SrTiO$_3$,
$|z_i| = 2$,
and a thickness of $t = 3.2$~nm,
one obtains $\Delta T_{cr} \approx \approx -115$~K.

\begin{figure}
\centering
\includegraphics[width=3.0in]{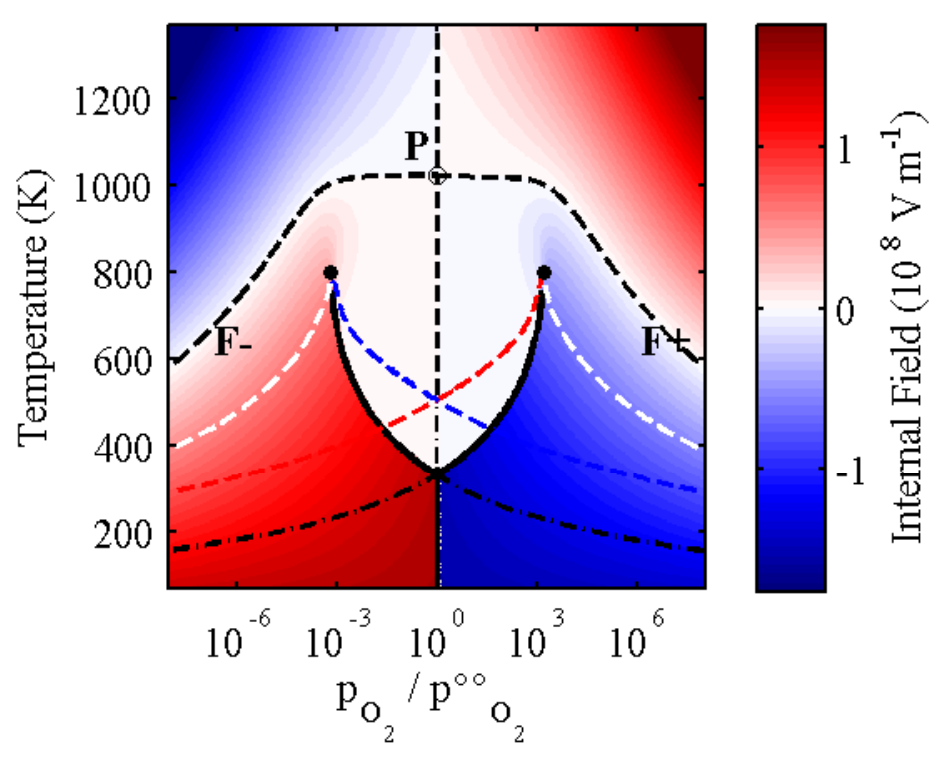} 
\caption{(Color online) Internal field for $t = 1.6$~nm,
corresponding to Fig.~\ref{F19a}(b).
Color scale gives electric field in stable phase.
Dashed black curves show conditions for zero field,
which intersect at $T_C^{\circ}$.}
%(chem\_ferroNTC3I\_22.pdf)}
\label{F20a}
\vspace{-0.1in}
\end{figure}

Figure~\ref{F20a} shows the internal field (along with the phase boundaries)
as a function of $p_{O_2}/p^{\circ\circ}_{O_2}$ and $T$ for a 1.6~nm thick film.
Like the electronic compensation cases,
Figs.~\ref{F2f} and~\ref{F6e},
the internal electric field is inverted in the polar phases near the phase boundaries because of the
incompletely neutralized depolarizing field.
While the electronic compensation model requires a nonzero screening length $\lambda$ 
to produce an inverted field,
the ionic compensation model does not.
The magnitude of the inverted field at the phase boundary
is much larger for ionic than for electronic compensation in the cases shown.
In all cases the inverted field regions extend above the critical point(s).
The oxygen partial pressure corresponding to zero internal field
can be obtained by setting the numerator in Eq.~(\ref{eq30}) to zero, giving
\begin{eqnarray}
\ln p_{O_2}^{E_{in}=0} &\approx& 
-\frac{2 n_i z_i e \lambda^{\dagger} P_0}{\epsilon_0 kT}
+ n_i \ln \left ( \frac{-A_i P_0/ z_i e}{1 + A_i P_0/ z_i e} \right )  \nonumber \\
&+& \frac{n_i \Delta G_i^{\circ\circ}}{kT},
\label{eq42}
\end{eqnarray}
where $P_0$ is
the $T$-dependent zero-field spontaneous polarization of the epitaxially strained film
given by the solution to Eq. (\ref{eq4}) with $E_{in} = 0$,
and $i = \om$ or $\op$ for positive or negative values of $P_0$.
As for electronic compensation,
the conditions for zero field are independent of film thickness,
and they intersect at $T_C^{\circ}$.
Rough values of the oxygen partial pressure at the critical points $p_{O_2}^{cr}$ 
can be obtained by assuming that the field is zero and neglecting the
first two terms in Eq.~(\ref{eq42}).
This gives $kT_{cr} \ln p_{O_2}^{cr} \approx \approx n_i \Delta G_i^{\circ\circ}$,
where $i = \om$ for positive $P$ (high $p_{O_2}$) and
$i = \op$ for negative $P$ (low $p_{O_2}$).

% figure 20a: Field vs T and pO2 for 1.6 nm film

\subsection{Phase diagram for controlled surface oxygen density}

\begin{figure}
\centering
\includegraphics[width=3.0in]{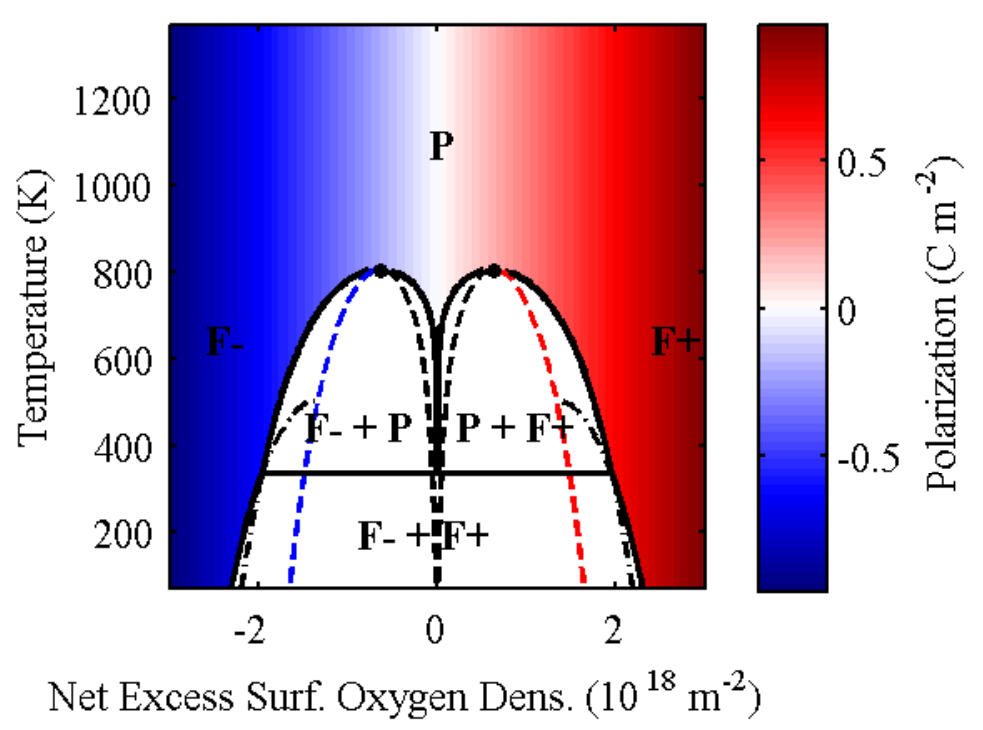} 
\caption{(Color online) 
Phase diagram
as a function of net excess surface oxygen density $\rho_O$ and $T$ for 1.6 nm of PbTiO$_3$ 
coherently strained to SrTiO$_3$,
corresponding to Fig.~\ref{F19a}(b).
Color scale gives polarization in single-phase region;
three two-phase regions are marked.
Solid and dash-dot black curves are stable and metastable phase boundaries,
respectively, between nonpolar paraelectric (P)
and positive and negative polar ferroelectric (F+ and F-) phases. 
Dashed red, blue, and black curves are metastability limits of the F+, F-, and P phases, 
respectively.}
%(chem\_ferroNTC3L\_21.pdf)}
\label{F22}
\vspace{-0.1in}
\end{figure}

\begin{figure}
\centering
\includegraphics[width=3.0in]{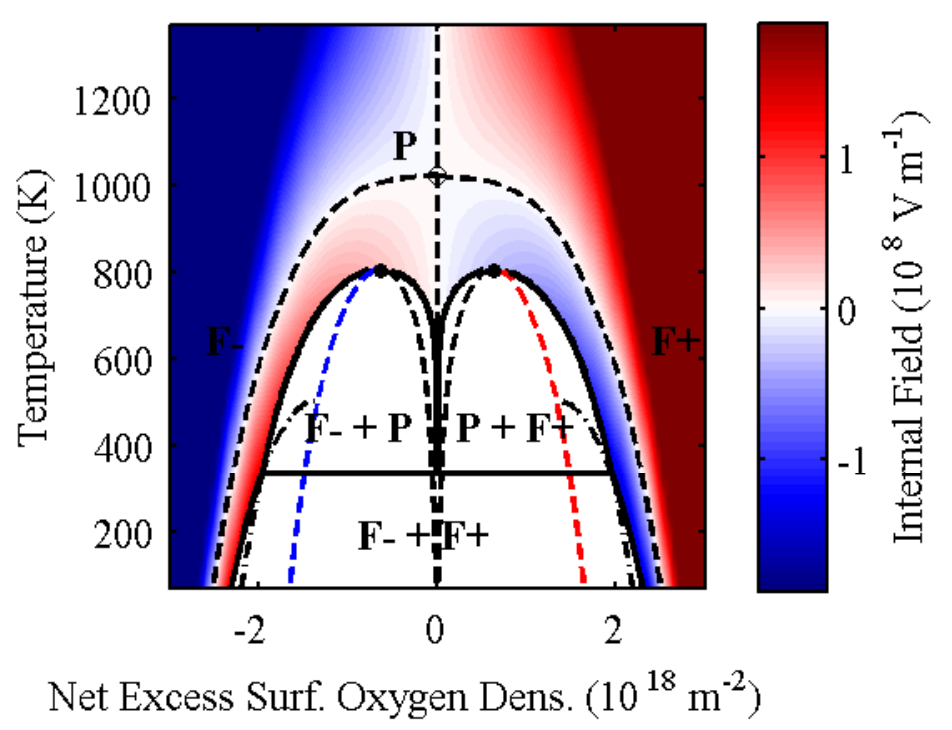} 
\caption{(Color online) Internal field
corresponding to Fig.~\ref{F22}.
Color scale gives electric field in stable phase.
Dashed black curves show conditions for zero field,
which intersect at $T_C^{\circ}$ (open circle).}
%(chem\_ferroNTC3L\_27.pdf)}
\label{F23}
\vspace{-0.1in}
\end{figure}

It is instructive to plot equilibrium phase diagrams for ionic compensation
as a function of the net excess surface oxygen density,
defined by
\begin{equation}
\rho_O \equiv \sum_i \frac{2 \theta_i}{n_i A_i}.
\end{equation}
In the typical case where we can neglect one or the other of the $\theta_i$
for positive or negative $\sigma$,
one obtains $\rho_O \approx 2 \sigma / n_i z_i e$,
for $i = \om$ or $\op$,
so that $\rho_O$ is simply proportional to $\sigma$ for each range.
Figures~\ref{F22} and \ref{F23} show the polarization and internal field plotted
as a color scale on 
$\rho_O$ vs. $T$ axes for $t =  1.6$ nm.
These correspond to the $p_{O_2}$ vs. $T$ diagrams
shown in Figs.~\ref{F19a}(b) and \ref{F20a}.
Such controlled $\rho_O$ diagrams can be obtained using the
Helmholtz free energy
\begin{eqnarray}
\label{eq43}
&{\cal A}& = {\cal G} + \frac{\rho_O} {2} kT \ln p_{O_2}
\approx  t  f(\sigma)
+ \frac{\lambda^{\dagger} \sigma^2}{\epsilon_0}
+ \frac{\sigma  \Delta G_i^{\circ\circ}}{z_i e}\\
&+& \frac{kT}{A_i} \left [ \frac{A_i \sigma}{z_i e} \ln \left ( \frac{A_i \sigma}{z_i e} \right )
+ \left ( 1- \frac{A_i \sigma}{z_i e} \right ) \ln \left ( 1 - \frac{A_i \sigma}{z_i e} \right ) \right ], 
\nonumber
\end{eqnarray}
where $i = \om$ or $\op$ as in Eq.~\ref{eq22}.

Because $\rho_O$ is a conserved order parameter,
the phase boundaries occuring on the controlled potential diagrams
become two-phase fields on Figs.~\ref{F22} and \ref{F23},
as in the electronic compensation case for controlled
surface charge, Figs.~\ref{F6d} and \ref{F6e}.
These show the phase separation that would occur
in a closed system if the net amount of excess surface oxygen
is fixed, rather than the external $p_{O_2}$.
The phase diagram for ionic compensation is more complex than that for
electronic compensation;
there are two equilibrium two-phase fields between
polar and nonpolar phases above a tie-line at the triple point temperature.

% figure 22: Polarization vs T and rho_O for 1.6 nm film

% figure 23: Field vs T and rho_O for 1.6 nm film

\section{Discussion}

The new parameters in the model developed above for ionic surface compensation are
$n_i$, $z_i$, $A_i$, and $\Delta G_i^{\circ\circ}$ for $i = \om$ and $\op$, 
as well as $\lambda^{\prime}$.
These can be related to the locations of the features on the phase diagram.
Approximate expressions are given above that show how the $\Delta G_i^{\circ\circ}$
determine $p^{\circ\circ}_{O_2}$ and the $p_{O_2}^{cr}$,
which give the center and width of the phase diagram in $p_{O_2}$ coordinates.
These expressions are particularly simple when the phase diagram
is symmetric in coordinates of
$p_{O_2}/p^{\circ\circ}_{O_2}$ versus $T$,
i.e. when $n_\op = - n_\om$, $z_\om = - z_\op$, and $A_\om = A_\op$.
In this case one obtains $\ln p^{\circ\circ}_{O_2} = n_\om(\Delta G_\om^{\circ\circ} - \Delta G_\op^{\circ\circ})/(2 k T)$,
$\ln (p_{O_2}^{cr}/p^{\circ\circ}_{O_2}) \approx \approx \pm n_\om(\Delta G_\om^{\circ\circ} + \Delta G_\op^{\circ\circ})/(2 k T_{cr})$.
The value of $\lambda^{\prime}$ affects
the critical temperatures $T_{cr}$,
which may be suppressed or enhanced relative to the electronic compensation value.

\begin{figure}
\centering
\includegraphics[width=2.5in]{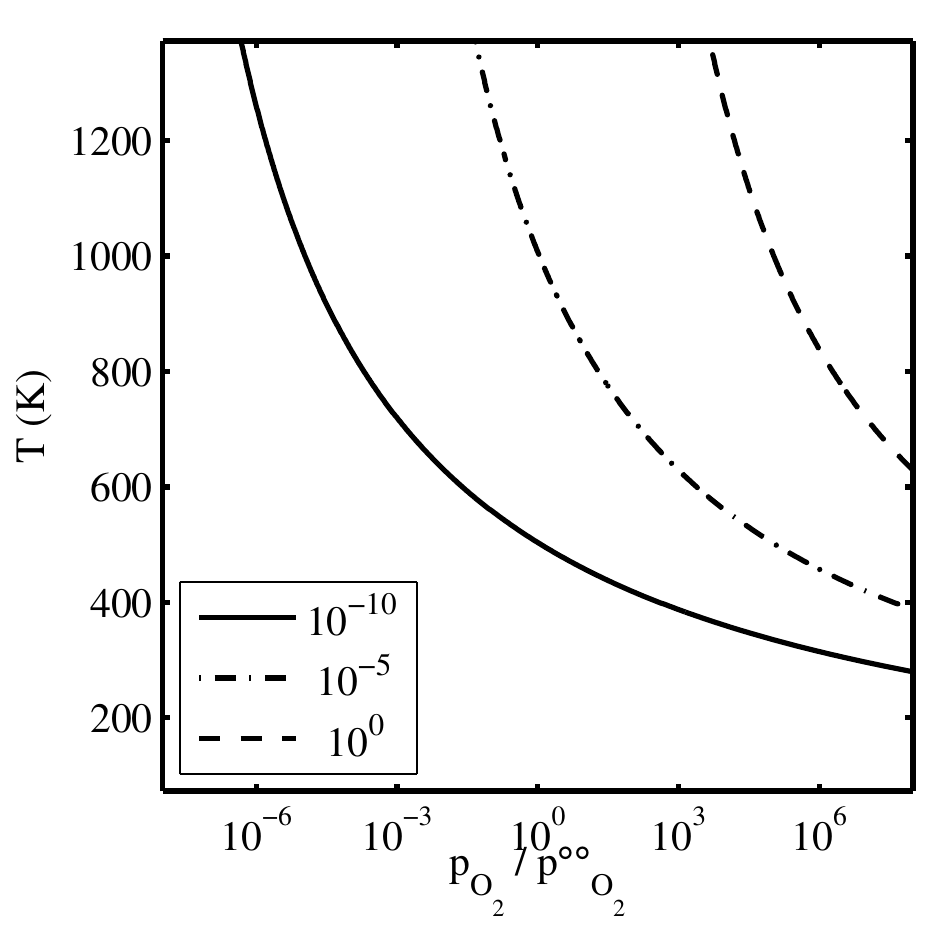} 
\caption{Curves show fixed values of $p_{O_2}$ (bar) 
given in the legend, plotted using same normalized axes used for the phase diagrams,
with parameters corresponding to Table~\ref{tab2}.}
%(chem\_ferroNTC3I\_12.pdf)}
\label{F18c}
\vspace{-0.1in}
\end{figure}

Although the phase diagrams we have shown are symmetric when plotted in
$p_{O_2}/p^{\circ\circ}_{O_2}$ versus $T$ coordinates,
the value of $p^{\circ\circ}_{O_2}$ is expected to be a function of $T$.
Thus experimental phase diagrams obtained as a function of
$p_{O_2}$ versus $T$ are not expected to be symmetric.
Figure~\ref{F18c} shows the trajectories of constant $p_{O_2}$
on the same $T$ vs.
$p_{O_2}/p^{\circ\circ}_{O_2}$ axes used to plot the phase diagrams.
Note that we have neglected any temperature dependence of $\Delta G_i^{\circ\circ}$
in these calculations.
Such temperature dependence might be expected
since the entropy of O$_2$ in the environment may be different than that of the adsorbed ions.

% figure 18c: pO2/pO2star(T)

While the Gibbs free energy expressions for the ionic and electronic compensation cases
are very similar,
the phase diagrams differ qualitatively because a different parameter is fixed.
If we neglect the polarization dependence of the $\Delta G_i^{\circ}$ so that $\lambda^{\prime} = 0$
and we consider $\theta_i << 1$ for both $i = \om$ and $\op$,
then by using the mass action equilibria Eqs.~({\ref{eq18})
one can show that the Gibbs free energy for ionic compensation, Eq.~({\ref{eq20a}),
reduces to Eq.~({\ref{eq5c}) used for the fixed $V_{ex}$ case.
However,
the fixed $V_{ex}$ and fixed $p_{O_2}$ conditions lead to different
equilibrium free energy surfaces, Figs.~\ref{F2aa} and~\ref{F10d},
even in the case of $\lambda^{\prime} = 0$ and $\theta_i << 1$.
The relationship between the fixed $V_{ex}$ and fixed $p_{O_2}$ conditions can be seen
from Eq.~(\ref{eq21}).
Here $p_{O_2}$ enters into the expression for $V_{ex}$   
simply through the term $kT \ln p_{O_2}/(z_i n_i e)$. 
In general, however, fixed $p_{O_2}$ does not correspond to fixed $V_{ex}$ because 
there are other terms and they depend upon polarization.
In particular, if the value of $\Delta G_i^{\circ\circ}/(z_i e)$ differs for positive and negative surface ions,
then there is an abrupt jump in $V_{ex}$ when crossing $P=0$. 
Thus fixed $V_{ex}$ and fixed $p_{O_2}$ constraints produce different equilibrium behavior
even when the free energy expression for both cases is the same.

The form of the phase diagrams in Fig.~\ref{F19a} in which a stable nonpolar phase intervenes
between the polar phases is due to the appearance of a third local minimum in ${\cal G}(\sigma)$ 
near $\sigma = 0$ as shown in Fig.~\ref{F10e}.
This is conceptually similar to the behavior of a ferroelectric with a first-order transition,
for which the coefficient of $P^4$ in the free energy expression is negative.~\cite{Strukov98,MERZ53PR}
For example, Figs.~\ref{F21} and \ref{F21a} in Appendix~\ref{app:firstorder}
show the phase diagrams for controlled internal field $E_{in}$ and net surface charge $\bar \sigma$
for unstressed bulk PbTiO$_3$  with ideal electrodes ($\lambda = 0$).
Here the topology of the equilibrium phase boundaries is similar to that in
Figs.~\ref{F19a}(a,b) and \ref{F22}, with a triple point and two critical points.
However, the range of temperatures spanned by this structure in ultrathin films
with ionic compensation can be much larger than in bulk PbTiO$_3$.
Furthermore, the polarization of the paraelectric phase
at temperatures below the critical points
is much closer to zero in ultrathin films with ionic compensation,
because of the sharp minimum in ${\cal G}(\sigma)$ at $\sigma = 0$.

The appearance of a stable nonpolar state
between the polar states on the $p_{O_2}$ vs. $T$ phase diagram
can affect the mechanism of switching
and the internal field at which switching occurs (i.e. the coercive field).
During switching by ramping $p_{O_2}$, 
the film may first become unstable with respect to the nonpolar state 
before reaching $p_{O_2}$ values that stabilize the opposite polarization,
thus suppressing nucleation of oppositely polarized domains.
In this case the internal field could reach the intrinsic coercive field
and switching occur by a continuous, spinodal mechanism without nucleation.
This could produce the recently observed crossover to a continuous mechanism\cite{10_HighlandPRL}
through an equilibrium pathway not requiring kinetic suppression of nucleation.

The model developed here contains several assumptions 
that could be relaxed in future extensions.
We assume that the effective screening length $\lambda$ is not negative,
so that electronic interfacial effects tend to suppress rather than enhance polarization
in ultrathin films.
We also neglect any polarization dependence of $\lambda$.
{\it Ab initio} calculations\cite{09_Stengel_NatMat,09_StengelPRB_80_224110} 
indicate that in some systems the interfaces enhance film polarization,
which can be modeled with a negative $\lambda$,
and that $\lambda$ depends on $P$.
These effects could be included by modifications to our electrostatic boundary conditions
and free energy expressions.
We constrain the free and bound charge at each interface to reside in single planes, 
so that there is no space charge.
Such space charge could be included as has been done previously in models 
with semiconducting ferroelectric films and/or 
electrodes.\cite{BAT73PRL, BAT73JVST,WURF76FERRO,BRAT00PRB}
Since the screening layer of thickness $\lambda$ is a conceptual construct 
rather than an actual dielectric layer in our model,
we do not consider tunneling of free charge across this layer,
which has been recently considered for systems with a dielectric separating the electrode
from the ferroelectric.\cite{JIANG09PRB}
These effects could be added for such systems.
We also neglect the possibility of equilibrium 180$^{\circ}$ stripe domain 
formation\cite{02_StriefferPRL_89_067601,04_Fong_Science,GBS06JAP,LAI07APL,CT08APL,BRAT08JCTN}
in which nanoscale domain structures reduce the depolarizing field 
even when there is little or no
electronic or ionic compensation charge at one or both interfaces.
A full treatment of equilibrium stripe domains for the ionic compensation case
would be valuable in future work.

\section{Summary and Conclusions}

Ionic compensation of a ferroelectric surface
due to chemical equilibrium with an environment
introduces new features into the phase diagrams, Figs.~\ref{F19a} and~\ref{F22},
not present in the standard phase diagrams for a second-order transition in a film
with electronic compensation, Figs.~\ref{F2e} and~\ref{F6d}.
The constant $p_{O_2}$ chemical boundary condition shown in Fig.~\ref{F7} is a hybrid between the constant
$V_{ex}$ and constant $\sigma$ boundary conditions shown in Fig.~\ref{F2}.
Because the surface concentrations of ionic species $\theta_i$ are limited to values
between zero and unity, constant surface charge regimes occur
when the $\theta_i$ are saturated.
In the regimes where one of the $\theta_i$ is varying between these limits, 
the boundary condition is similar to a fixed $V_{ex}$ condition.
There are two independent relations for the surface charge $\sigma$ as a function of
$p_{O_2}$, depending upon whether positive or negative surface ions predominate.
In the $p_{O_2}$ region where there is insufficient surface charge of either sign 
to stabilize a polar state,
the nonpolar state becomes stable between the positive and negative polar states,
producing two critical points, a triple point,
and a strong dependence of $T_C$ on $p_{O_2}$.
Large inverted internal fields occur at equilibrium in the polar phases
near the phase boundaries with the nonpolar phase.
Manipulation of ultrathin ferroelectric films via controlled ionic compensation 
may thus allow experimental access to
exotic nonpolar and high-field states such as those modeled in recent 
{\it ab initio} calculations\cite{09_StengelNatPhys_5_304,BECKMAN09PRB}
that would not be stable under electronic compensation conditions.

\section*{ACKNOWLEDGMENTS}

We have benefited greatly from discussions with and experimental results obtained by 
our collaborators T.~T. Fister, M.-I. Richard, D.~D.~Fong, P.~H.~Fuoss, C.~Thompson, 
J.~A.~Eastman, and S.~K.~Streiffer,
as well as comments from M. Stengel and D. Vanderbilt.
Work supported by the U.S. Department of Energy, Office of Science, 
Office of Basic Energy Sciences, Division of Materials Sciences and Engineering,
under Contract DE-AC02-06CH11357.  

%\vfill
%\eject
\appendix

\section{Phase Diagrams for First-Order Ferroelectric}
\label{app:firstorder}

The phase diagrams for a bulk ferroelectric that has a first-order transition at zero field
are similar to those for ultrathin films with ionic surface compensation.
Here we present calculated phase diagrams showing the region near the critical points
for the weakly first-order transition in unstressed bulk PbTiO$_3$.
The features in these phase diagrams can be compared with
those for ultrathin epitaxially strained films shown above,
which have a second-order transition for electronic compensation
but a strongly first-order transition for ionic surface compensation.

% figure 21: Polarization phase diagram for unstressed PTO as a function of field and T
\begin{figure}
\centering
\includegraphics[width=3.0in]{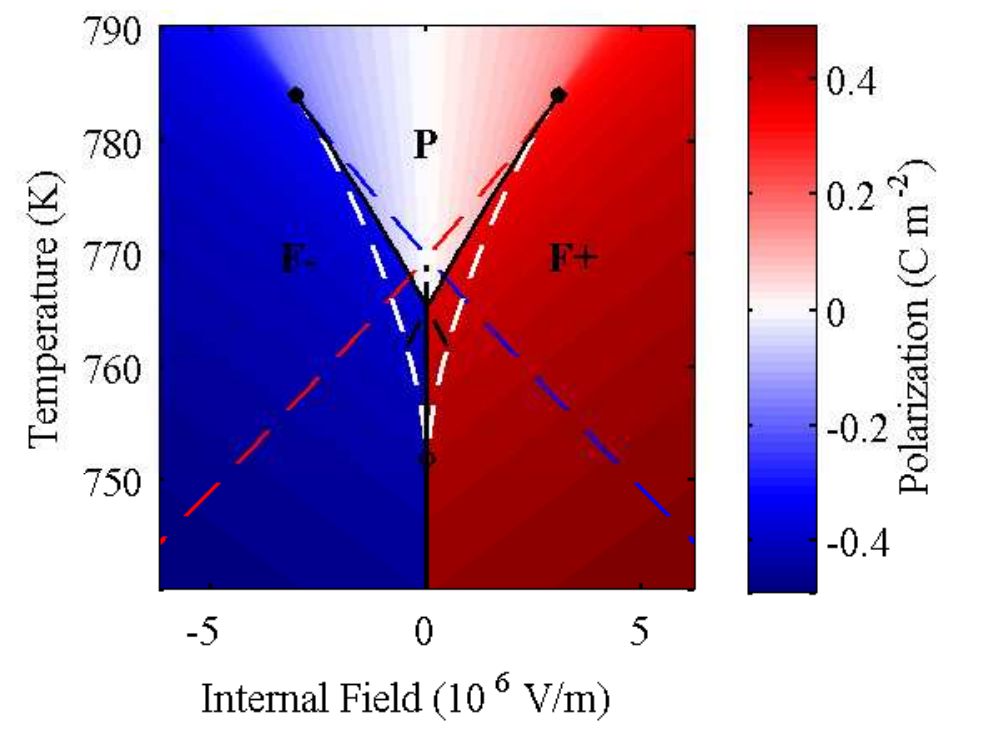} 
\caption{(Color online) Region near critical points on equilibrium polarization phase diagram 
as a function of $E_{in}$ and $T$,
for unstressed bulk PbTiO$_3$ with ideal electrodes ($\lambda = 0$).
Color scale gives polarization of stable phase.
Solid and dash-dot black curves are stable and metastable phase boundaries,
respectively, between nonpolar paraelectric (P)
and positive and negative polar ferroelectric (F+ and F-) phases. 
Dashed red, blue, and white curves are metastability limits of the F+, F-, and P phases, 
respectively.}
%(chem\_ferro\_firstorder\_3.pdf)}
\label{F21}
\end{figure}

% figure 21a: Polarization phase diagram for unstressed PTO as a function of $\sigma$ and T
\begin{figure}
\centering
\includegraphics[width=3.0in]{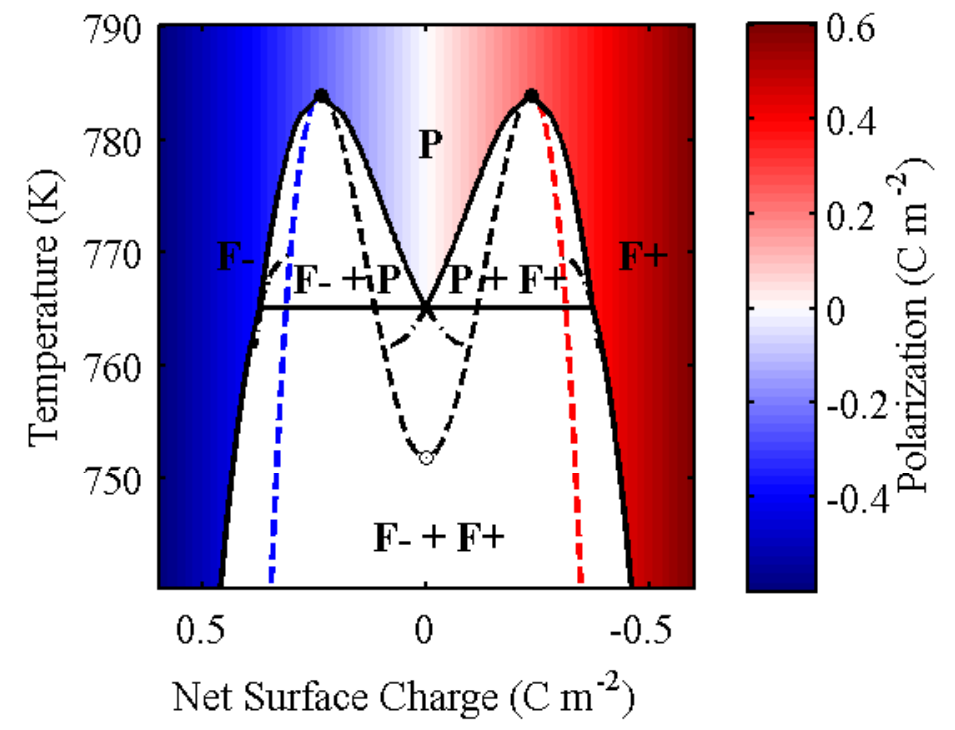} 
\caption{(Color online) Region near critical points on equilibrium polarization phase diagram
for unstressed bulk PbTiO$_3$ with ideal electrodes ($\lambda = 0$)
corresponding to Fig.~\ref{F21},
here plotted as a function of net surface charge ${\bar \sigma}$  and $T$.
Color scale gives polarization in single-phase region,
which is separated by phase boundary (solid black curve) from three two-phase fields;
metastable phase boundaries are dash-dot curves, and
two critical points are marked by filled circles. 
Dashed red, blue, and black curves are metastability limits 
of the F+, F-, and P phases, respectively.}
%(chem\_ferro\_firstorder\_4.pdf)}
\label{F21a}
\end{figure}

These phase diagrams are calculated using the Landau-Ginzburg-Devonshire expression for the free energy per unit volume
of unstressed bulk PbTiO$_3$ with ideal electrodes having zero screening length ($\lambda = 0$),
\begin{equation}
{\cal G}_v = \alpha_{1}P^2 + \alpha_{11}P^4 + \alpha_{111}P^6 + \epsilon_0 E_{in}^2/2 - E_{in}P
\end{equation}
with $\alpha_{1} = (T-T_0)/2 \epsilon_0 C$,
where the parameters are given in Table I of the main paper.
Figure~\ref{F21} shows the phase diagram for controlled internal field $E_{in}$
and Fig.~\ref{F21a} shows the phase diagram for  controlled net surface charge $\bar \sigma$.
These are typical for a ferroelectric with a first-order transition,
for which the coefficient of $P^4$ in the free energy expression is negative.
The topology of the equilibrium phase boundaries is similar to the ionic compensation case,
Figs.~\ref{F19a}(a,b) and \ref{F22}, with a triple point and two critical points.
However, the paraelectric phase above the triple point has a much wider range of polarization around zero
in this case,
reflecting the relatively broad minimum in ${\cal G}_v$ 
near $P = 0$.

%\eject

\end{document}